\documentclass[a4paper,11pt]{article}
\usepackage{jheppub} 
\usepackage{hyperref}
\usepackage{amsmath}

\usepackage[most]{tcolorbox}
\usepackage[normalem]{ulem}

\usepackage{tikz}
\usetikzlibrary{shapes}
\usetikzlibrary{positioning}
\usetikzlibrary{fit, backgrounds, shadows, shadows.blur}

\usepackage{tkz-euclide}
\definecolor{boxblue}{RGB}{0,153,204} 

\newtcolorbox{conjectureBox}[1][]{
  breakable,
  colback=white,
  colframe=black,
  title={Conjecture},
  fonttitle=\bfseries,
  #1
}

\newcounter{mythm}

\newtcolorbox{observationBox}[1][]{
  breakable,
  colback=white,
  colframe=blue,
  fonttitle=\bfseries,
  title=Observation~\refstepcounter{mythm}\themythm,
  #1
}

\newtcolorbox{proofBox}[1][]{
  breakable,
  colback=white,
  colframe=blue,
  title=Proof,
  fonttitle=\bfseries,
  #1
}

\definecolor{green}{rgb}{0.3,0.7,0.3}
\newcommand{\Blue}{\color{black}}

\newcommand{\Zred}{Z_{d-2}}
\newcommand{\Sred}{S_{d-2}}
\newcommand{\thetab}{{\bar \theta}}
\newcommand{\x}{{\texttt{x}}}
\newcommand{\y}{{\texttt{y}}}
\newcommand{\ii}{{\texttt{i}}}
\newcommand{\jj}{{\texttt{j}}}

\newcommand{\psib}{{\bar \psi}}

\newcommand{\Ocal}{\mathcal{O}}
\definecolor{boxdef}{RGB}{210,238,227}
\newcommand{\mycolor}{teal!25!white}

\newcommand\myatop[2]{\genfrac{}{}{0pt}{}{#1}{#2}}

\title{\centering Unleash $Q$! \\
Cohomology, Localization,  and Interpolation\\
in Parisi--Sourlas Supersymmetry}

\hypersetup{pdftitle = {Unleash Q! Cohomology, Localization,  and Interpolation in Parisi-Sourlas Supersymmetry}}

\author[\ensuremath{\mathcal{Q}},\ensuremath{\mathfrak{Q}}]{Bruno Le Floch}
\author[\ensuremath{\mathcal{Q}},Q]{Gela Patashuri}
\author[\ensuremath{\mathcal{Q}},\ensuremath{\mathfrak{Q}},\ensuremath{\mathbb{Q}}]{Emilio Trevisani}
\hypersetup{pdfauthor = {Bruno Le Floch, Gela Patashuri, Emilio Trevisani}}

\affiliation[\ensuremath{\mathcal{Q}}]{Laboratoire de Physique Théorique et Hautes Energies,
Sorbonne Université, 4 Place Jussieu, 75252 Paris, France}
\affiliation[\ensuremath{\mathfrak{Q}}]{Centre National de la Recherche Scientifique, Paris, France}
\affiliation[Q]{Department of Physics,
The Ohio State University,
Columbus, OH 43210, USA}
\affiliation[\ensuremath{\mathbb{Q}}]{Department of Theoretical Physics, CERN, 1211 Meyrin, Switzerland}

\emailAdd{blefloch@lpthe.jussieu.fr}\emailAdd{patashuri.1@osu.edu}\emailAdd{trevisani@lpthe.jussieu.fr}

\abstract{
Parisi--Sourlas supersymmetric models are known to undergo dimensional
reduction; that is, their physics is captured by models in two fewer dimensions.
In this work, we revisit dimensional reduction, providing new arguments
and reformulating existing proofs in terms of the cohomology of a
supercharge $Q$.
We obtain three main results.  First, we show that the recently developed picture of dimensional reduction via decoupling of operators admits a natural explanation in terms of $Q$-exactness.  Second, we provide a new proof of dimensional reduction using the supersymmetric localization argument. Third, we revisit Cardy’s
``interpolation'' proof, which is reminiscent of localization but does
not rely on saddle-point methods, and show that it can be understood as a consequence of deforming the action by a $Q$-exact term. Finally, we show that, while existing nonperturbative proofs of dimensional reduction
apply only to scalar Lagrangians, our formulation of Cardy’s argument extends to any
theory with Parisi--Sourlas supersymmetry.
}

\begin{document}
\maketitle
\flushbottom

\section{Introduction}
\label{sec:intro}
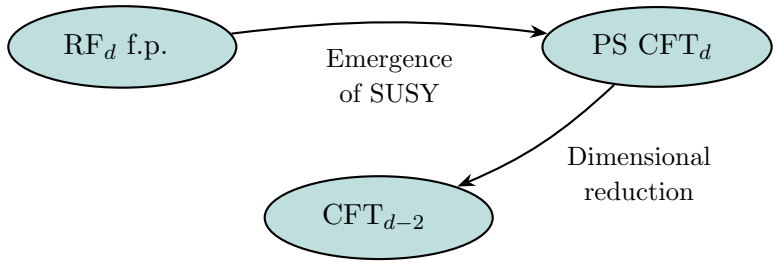
\begin{figure}[h!]
\centering
\begin{tikzpicture}[
  >=Stealth,
  every node/.style={font=\normalsize},
  oval/.style={
    ellipse,
    draw=black,
    thick,
    fill= \mycolor,
    align=center,
    inner sep=6pt
  }
]
\node[oval, minimum width=3.0cm, minimum height=1.0cm] (rf)
  {RF$_{d}$ f.p.
  };

\node[oval, minimum width=3.0cm, minimum height=1.0cm, right=4cm of rf] (pscft)
  {PS CFT$_{d}$
  };

\node[oval, minimum width=3.0cm, minimum height=1.0cm, below=1.5cm of $(rf)!0.65!(pscft)$,
  xshift=-12mm, yshift=-2mm] (cft)
  {CFT$_{d-2}$ 
  };

\draw[->, thick]
  (rf) to[bend left=7pt]
  node[midway, below=6pt, font=\small, align=center] {Emergence \\ of SUSY}
  (pscft);

\draw[->, thick]
  (pscft) to[bend left=10pt]
  node[midway, below, right,  
  font=\small, align=center] {}
  (cft);

\node[font=\small] at (6.8, -1.7)  (c)     {
 {\begin{tabular}{c}
    Dimensional \\ reduction
\end{tabular}}
};

\end{tikzpicture}
\caption{
\label{fig:triangle}
Diagram showing the relations between the fixed point of a random field theory in $d$ dimensions ($\mathrm{RF}_{d}$ f.p.), a PS $\mathrm{CFT}_{d}$, and a pure $\mathrm{CFT}_{d-2}$. }
\end{figure}

Parisi--Sourlas (PS) supersymmetry (SUSY) plays an important role in statistical physics in explaining the behavior of random field (RF) models at criticality.
Historically, it was first argued in~\cite{PhysRevLett.37.1364} that random field models undergo dimensional reduction, i.e., their infrared properties are captured by a model in two fewer dimensions (see
also~\cite{Fytas_2019, PhysRevE.95.042117, PhysRevLett.122.240603, PhysRevLett.116.227201,
PhysRevLett.46.871, PhysRevLett.35.1399,CARDY1985123,Brezin_1998, PhysRevLett.88.177202, PhysRevB.78.024203,
Tarjus:2024mop, Kaviraj:2020pwv, Kaviraj:2021qii, Kaviraj:2022bvd, Rychkov:2023rgq, Piazza:2024wll}
for more works in this direction).
This unconventional phenomenon was later understood as a consequence of the emergence of PS SUSY~\cite{Parisi:1979ka} close to the IR fixed point, according to the logic of Fig.~\ref{fig:triangle}. 
The emergence of PS supersymmetry depends sensitively on the RF model and the spacetime dimensions. For instance, it occurs in the RF Ising model in five dimensions but not in 
$d=3,4$, whereas in the RF $\phi^3$ model it is realized for all $2 \leq d<8$.
In contrast, dimensional reduction in models with PS SUSY is expected to work \textit{in all cases}, including for QFTs away from their IR fixed point.

The dimensional reduction of scalar actions invariant under PS SUSY was proven in several ways over the decades \cite{Parisi:1979ka,KLEIN1983473, Klein:1984ff,CARDY1983470, Zaboronsky:1996qn, paperI, Ghosh:2025lzb}.
It was originally understood only perturbatively~\cite{Parisi:1979ka},  by noticing that each Feynman diagram could be rewritten as a lower-dimensional one.
Later, Cardy found a beautiful non-perturbative argument, which is reminiscent of localization but does not invoke any saddle-point limit~\cite{CARDY1983470}. This argument uses a deformed model which,  by tuning a parameter, interpolates between the SUSY and reduced actions; the model is then shown to be independent of the interpolation parameter.

A very early localization argument was considered by Zaboronsky~\cite{Zaboronsky:1996qn}, who used a non-standard approach where superfield configurations are 
localized directly to fixed points of a supercharge in superspace, rather than the space of fields.
This predates the wave of supersymmetric localization results spearheaded by Pestun~\cite{Pestun:2007rz} (see the review~\cite{Pestun:2016zxk} and references therein).

More recently, in \cite{paperI}, a geometric superspace argument was found that showed that, when restricted to $d-2$ dimensions, infinite towers of operators of the PS theory decouple and that the only non-trivial correlators of operators are in one-to-one correspondence with those of the lower-dimensional theory. 
In the same work, an axiomatic CFT explanation of the phenomenon was also given. This set the basis for the so-called \textit{dimensional uplift} of CFTs, in which a given CFT can be uplifted to two higher dimensions at the price of introducing PS supersymmetry; this was explored at the level of correlation functions of local and extended operators~\cite{paperI, Trevisani:2024djr, vanVliet:2025swv, Ghosh:2025lzb}.
See also~\cite{PARISI1982321, Brezin:1984ie, Kaviraj:2024cwf, 05b5fc40-d3a0-33cf-b199-5d8140f70420, Cardy:2023zna, Nakayama:2024jwq}
for other applications of PS SUSY\@.

The main aim of this paper is to rephrase old proofs and find new proofs of dimensional reduction using a unifying motif: \textit{given a supercharge $Q$ of the PS theory, $Q$-exact deformations should decouple}. 
This standard logic for high-energy physics is, in fact, not made explicit in most dimensional reduction arguments. 
To this end, we will provide three complementary  arguments: 
\begin{itemize}
    \item \underline{\textit{Cohomology argument}}: The cohomology argument is a rephrasing of the geometric decoupling argument of~\cite{paperI} in terms of the cohomology of~$Q$. In particular, we show that the towers of decoupled operators of~\cite{paperI} are $Q$-exact, while the dimensionally reduced operators are defined by the cohomology of~$Q$. This provides a bridge between~\cite{paperI} and the cohomological construction in~\cite{Beem:2013sza}, 
    which finds that a protected sector of operators in any four-dimensional $\mathcal{N}=2$ superconformal field theory organizes into a two-dimensional chiral CFT (see also e.g.,~\cite{Beem:2014kka, Beem:2014rza, Lemos:2014lua, Chester:2014mea, Dedushenko:2016jxl, Dedushenko:2019mnd} for extensions to other dimensions).
    \item \underline{\textit{Localization argument}}: The localization argument is original to this work and consists of adding a canonical $Q$-exact term to the action and following the standard SUSY localization procedure. 
    In addition, we show how this differs from the earlier work of Zaboronsky \cite{Zaboronsky:1996qn}, who used the same supercharge $Q$ but a non-conventional $Q$-exact deformation term, which gives rise to various subtleties.
    \item \underline{\textit{Interpolation argument}}: The interpolation argument is the one originally proposed by Cardy in \cite{Cardy:1983aa}. We show that it can be rephrased by saying that the difference of the PS action and the dimensionally reduced one is $Q$-exact. This argument is very powerful and elegant, since it allows for a smooth transition between the two actions, without the need to compute a one-loop determinant.
    We further show that the interpolation argument can be easily generalized to other PS models, which also comprise defects, thus providing a new non-perturbative proof of dimensional reduction for~\cite{Ghosh:2025lzb}.
\end{itemize}
With this work, we thus uncover the hidden cohomological structure behind the proofs, making them logically clearer and less accidental, rather than a consequence of special properties of PS SUSY\@.
This also shows that some of the arguments of dimensional reduction ---like the interpolation argument by Cardy \cite{Cardy:1983aa} and the decoupling argument \cite{paperI}--- have the potential to be used in the \textit{standard SUSY realm}.
Notice that the interpolation argument allows one to localize a model without having to compute the one-loop determinant, which makes it a very promising tool for theories defined on non-compact spaces.
The decoupling argument is also very powerful, since it applies under very simple assumptions, which amount to finding, in superspace, a reduced manifold with a supertraceless orthogonal metric. This could be useful for classifying when a sector of a SUSY theory admits a lower-dimensional description.

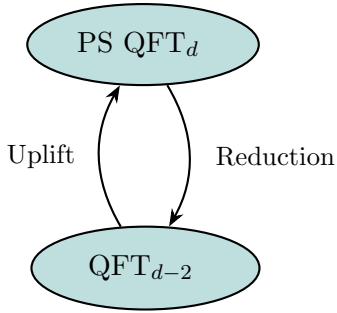
\begin{figure}
\centering
\begin{tikzpicture}[
  >=Stealth,
  every node/.style={font=\normalsize},
  oval/.style={
    ellipse,
    draw=black,
    thick,
    fill= \mycolor,
    align=center,
    inner sep=6pt
  }
]


\node[oval, minimum width=3.0cm, minimum height=1.0cm, right=0cm of rf] (pscft)
  {PS QFT$_{d}$
  };

\node[oval, minimum width=3.0cm, minimum height=1.0cm, below=2.2cm of $(rf)!0.65!(pscft)$,
  xshift=11mm, yshift=-2mm] (cft)
  {QFT$_{d-2}$ 
  };
\draw[->, thick]
  (cft) to[bend left]
  node[midway, left=5pt, font=\small] {Uplift}
  (pscft);

\draw[->, thick]
  (pscft) to[bend left]
  node[midway, right=6pt, font=\small] {Reduction}
  (cft);

\end{tikzpicture}
\caption{\label{fig:uplift}  PS uplift conjecture: any  $\mathrm{QFT}_{d-2}$ can be embedded into PS $\mathrm{QFT}_{d}$.}
\end{figure}

In this work, we also discover that Cardy’s interpolation argument can be generalized to prove dimensional reduction for any PS-supersymmetric action, not only scalar ones.
Since the types of PS and reduced actions can be quite generic, this suggests that, starting from any Lagrangian model in $d-2$ dimensions, one can construct a $d$-dimensional counterpart invariant under PS SUSY that reduces to it, as in Fig.~\ref{fig:uplift}.
It would be interesting to explore this in more detail and possibly rule out the existence of lower-dimensional models that do not have a higher-dimensional PS SUSY counterpart. 
This exploration would either support the conjecture of the existence of PS uplifts for any QFT of ~\cite{paperI, Trevisani:2024djr, vanVliet:2025swv, Ghosh:2025lzb}, or help identify the additional qualifications required for its validity.

The plan of the paper is as follows. In \autoref{sec:PS_SUSY}, we review PS supersymmetry and dimensional reduction.
We then discuss the specific supercharge~$Q$ that will play a cardinal role in the rest of the paper.
In \autoref{sec:cohomology}, we review the decoupling argument of~\cite{paperI}, and we rephrase it in terms of the cohomology of~$Q$.  
We then address localization arguments for dimensional reduction. 
In  \autoref{sec:Localization}, we discuss a new localization argument, and we show 
how this differs from Zaboronsky's argument of \cite{Zaboronsky:1996qn}. 
In \autoref{sec:Cardy}, we first review Cardy's interpolation argument of  \cite{Cardy:1983aa}, then show that the deformation is $Q$-exact, and finally explain how this can be applied to generic PS SUSY actions.
Finally, we end with some discussions in \autoref{sec:conclusions} and provide extra details in the appendices.

\section{Parisi--Sourlas supersymmetry and preliminaries}
\label{sec:PS_SUSY}
\subsection{Parisi--Sourlas symmetries and superspace}

The Parisi--Sourlas (PS) action, to be specified below, is invariant under \emph{scalar} supersymmetry transformations. In this section, we review the basic properties of the superfield formalism that make this supersymmetry manifest and set the conventions used throughout the paper.

We consider a $(d|2)$–dimensional superspace with coordinates
\begin{equation}
    y^a \equiv \bigl(x^\mu, \theta, \bar{\theta}\bigr) \, ,
\end{equation}
where $x^\mu$ ($\mu = 1,\ldots,d$) are commuting coordinates on a
$d$-dimensional Euclidean space, and $\theta$, $\bar{\theta}$ are
Lorentz-scalar fermionic (aka.~Grassmann) coordinates satisfying
\begin{equation}
    \theta^2 = 0\,, 
    \qquad \bar{\theta}^2 = 0\,, 
    \qquad \theta \bar{\theta} = - \bar{\theta} \theta\,.
\end{equation}
The super-index $a = (\mu, \theta, \bar{\theta})$ collects both bosonic and fermionic coordinates.

The graded commutation properties of the superspace coordinates can be
encoded in terms of their Grassmann parity $[a] \in \{0,1\}$, defined by
\begin{equation}
    [a] =
    \begin{cases}
        0, & a = \mu \, , \\
        1, & a = \theta, \bar{\theta}\, ,
    \end{cases}
    \label{eq:grassParity}
\end{equation}
so that the coordinates satisfy the graded commutation relation:
\begin{equation}
    y^a y^b = (-1)^{[a][b]} y^b y^a \, .
\end{equation}

To discuss supersymmetry transformations, it is convenient to equip the
superspace with a flat graded metric $g_{ab}$ whose only non-vanishing
components are
\begin{equation}
    g_{\mu\nu} = \delta_{\mu\nu}\,,
    \qquad
    g_{\theta\bar{\theta}} = - g_{\bar{\theta}\theta} = -1 \, .
\end{equation}
In the ordered basis $y^a = (x^\mu,\theta,\bar{\theta})$ this can be written as
\begin{equation}
    g_{ab} \equiv
    \begin{pmatrix}
        I_d & 0 \\
        0   & J_2
    \end{pmatrix},
    \qquad
    I_d \equiv \operatorname{diag}(\overbrace{1,\ldots,1}^{d})\, ,
    \qquad
    J_2 \equiv
    \begin{pmatrix}
        0 & -1 \\
        1 & 0
    \end{pmatrix} .
\end{equation}
The metric is graded-symmetric, i.e. $g_{ab} = (-1)^{[a][b]} g_{ba}$. The inverse metric $g^{ab}$ is defined by $g^{a b} g_{a c}=\delta^b_c$ and obeys $g^{b a} g_{c a}=\delta^b_c$; in particular, $g^{\mu\nu} = \delta^{\mu\nu}$ and
$g^{\bar{\theta}\theta} = - g^{\theta\bar{\theta}} = 1$.
We use $g_{ab}$ and $g^{ab}$ to lower and raise super-indices:
\begin{equation}
    y_a = g_{ab} y^b \, ,\qquad
    y^a = g^{ba} y_b \, ,
\end{equation}
that is, $y_a=(x_\mu,-\bar{\theta},\theta)$.
The corresponding superspace norm is
\begin{equation}
    y^2 \equiv y^a y_a
    = y^a g_{ab} y^b
    = x^2 - 2 \theta \bar{\theta}\,,
    \label{norm}
\end{equation}
where $x^2 \equiv x^\mu x_\mu = \delta_{\mu\nu} x^\mu x^\nu$.

Superspace derivatives are defined as
\begin{equation}
    \partial_a \equiv (\partial_\mu, \partial_\theta, \partial_{\bar\theta})
    \, ,
    \qquad
    \partial_\mu \equiv \frac{\partial}{\partial x^\mu}
    \,, 
    \quad
    \partial_\theta \equiv \frac{\partial}{\partial \theta} \, ,
    \quad
    \partial_{\bar\theta} \equiv \frac{\partial}{\partial \bar\theta} \, .
\end{equation}
Using the superspace metric to raise indices, $\partial^a \equiv g^{ba} \partial_b =(\partial^\mu,\partial_{\thetab},\partial_{\theta})$,
the super-Laplacian takes the form
\begin{equation}
    \partial^a \partial_a
    = \partial_x^2 + 2\,\partial_{\bar\theta}\partial_\theta \, ,
\end{equation}
where $\partial_x^2 \equiv \delta^{\mu\nu}\partial_\mu\partial_\nu$.
We normalize the integration over the Grassmann coordinates as follows 
\begin{equation}
     \int [d\bar\theta]\,[ d\theta] = \frac{1}{2\pi} \int d\bar\theta\, d\theta \, , \label{fermionic_measure}
\end{equation}
and we also use this notation for the fermionic measure in the path integral.

In superspace, PS SUSY can be simply understood as the set of transformations that preserve the superspace distance. We will be mostly interested in the following scalar PS SUSY action 
\begin{equation}\label{eq:PS_action_superfield} 
    S_{\text{SUSY}}= 2\pi\int d^{d|2}y\left[-\frac{1}{2} \Phi \partial^a \partial_a \Phi+V(\Phi)\right] \, ,
\end{equation}
where $d^{d|2}y \equiv d^d x [d \bar{\theta}] [d \theta]$ and the scalar superfield $\Phi$ can be  expanded in  terms of the components  $\varphi, \psi, \bar{\psi}$, $\omega$
as follows, 
\begin{equation}
    \Phi(x,\theta,\bar{\theta}) = \varphi(x) + \theta\bar{\psi}(x) + \bar{\theta}\psi(x) + \theta\bar{\theta}\omega(x) \, .
    \label{superscalar}
\end{equation}
Integrating out the anticommuting coordinates yields
\begin{equation}
\label{eq:PS_action}
    S_{\text{SUSY}}=\int 
    d^dx\Bigl( \partial^\mu \omega \partial_\mu \varphi-\omega^2+\omega V^{\prime}(\varphi)+\partial^\mu \psi \partial_\mu \bar{\psi}+\psi \bar{\psi} V^{\prime \prime}(\varphi) \Bigr) \, .
\end{equation}
Because of its superfield construction, the Parisi-Sourlas action is invariant under
$ ISO(d|2) $, that is, the super Poincaré symmetry group $\mathbb{R}^{d|2} \rtimes OSp(d|2)$ built out of supertranslations and superrotations 
which both preserve the norm \eqref{norm}.

Let us discuss the algebra of the generators. 
To do so we shall introduce the graded commutator $[X, Y\}$, defined as the commutator $[X, Y\}=[X, Y]$ if one or both of $X$ and $Y$ are Grassmann-even, and as the anticommutator $[X, Y\}=\{X, Y\}$ if both $X$ and $Y$ are Grassmann-odd.

The supertranslation generators simply graded-commute, namely $[P^a,P^b\}=0$.
Superrotation generators satisfy the orthosymplectic algebra of $OSp(d|2)$, 
\begin{equation}
\label{algebra_M_OSp}
 \!\!   [M^{ab}, M^{cd}\} = - g^{cb} M^{ad} + (-1)^{[a][b]} g^{ca} M^{bd} + (-1)^{[c][d]} g^{db} M^{ac} - (-1)^{[a][b] + [c][d]} g^{da} M^{bc} \, .
\end{equation}
Finally, supertranslations transform as vectors under superrotations; therefore, their algebra takes the form
\begin{equation}
    [M^{a b}, P^c\}=-g^{c b} P^a+(-1)^{[a][b]} g^{c a} P^b \,.
\end{equation}

It is useful to introduce the Killing vectors associated with the generators,
\begin{equation}
    p^a=\partial^a \, , \qquad m^{ab} = y^a \partial^b - (-1)^{[a][b]}y^b \partial^a \, ,
\end{equation}
which can be easily shown to satisfy the opposite commutation relations as $P^a$ and $M^{ab}$.
For more details on the charges and their algebra, we refer to appendix \ref{sec: App A}.

\subsection{Review of dimensional reduction}
\label{sec:dim_red}

The expectation value of superfield operators in the supersymmetric
theory is defined by the path integral
\begin{equation}
\langle \mathcal{O}(y) \ldots \rangle_{d|2}
\equiv
\frac{1}{Z_{\text{SUSY}}}
\int \mathcal{D}\Phi \,
\mathcal{O}(y) \ldots \,
e^{-S_{\text{SUSY}}[\Phi]} \, ,
\label{eq:expVal}
\end{equation}
where $Z_{\text{SUSY}}$ denotes the partition function,
$Z_{\text{SUSY}} = \int \mathcal{D}\Phi \, e^{-S_{\text{SUSY}}[\Phi]}$,
and the action $S_{\text{SUSY}}$ is defined in
\eqref{eq:PS_action_superfield}.

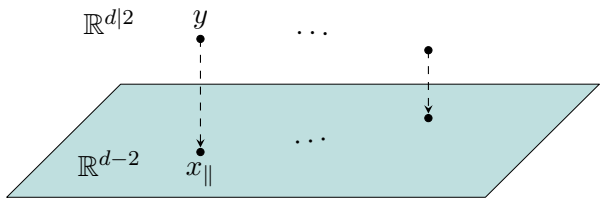
\begin{figure}[htbp]
\centering
\begin{tikzpicture}[scale=1.5]
\tkzInit[xmin=0,xmax=5.5,ymin=0,ymax=1.5] 
    \tkzClip[space=.5] 
    \tkzDefPoint(0.3,0){A} 
    \tkzDefPoint(4.5,0){B} 
    \tkzDefPoint(5.5,1){C} 
    \tkzDefPointWith[colinear= at C](B,A) \tkzGetPoint{D}
       \tkzDrawPolygon[fill= \mycolor](A,B,C,D)
    \tkzDefPoint(2. ,0.4){X1} 
      \tkzDefPoint(2. ,0.44){X1p} 
       \tkzDefPoint(2. ,1.4){Y1} 
     \tkzDefPoint(4. ,0.7){Xn} 
        \tkzDefPoint(4. ,0.74){Xnp} 
          \tkzDefPoint(4. ,1.3){Yn} 
          \draw[fill=black, draw=black] (X1) circle (0.9pt);
           \draw[fill=black, draw=black] (Xn) circle (0.9pt);
            \draw[fill=black, draw=black] (Y1) circle (0.9pt);
           \draw[fill=black, draw=black] (Yn) circle (0.9pt);
        \draw[dashed, -stealth](Y1)->(X1p);
       \draw[dashed, -stealth](Yn)->(Xnp);
    \node[below] at (X1) { $ x_\parallel$
    };
  \node[below] at (Xn) { 
  };
  \node[above] at (Y1) { $ y$};
  \node[above] at (Yn) {};
  \node[above,rotate=-2] at (3.,1.35) { $ \dots$};
  \node[above,rotate=7] at (3.,0.4) { $ \dots$};
   \node[above] at (1.2,1.35) {$\mathbb{R}^{d\mid 2}$};
    \begin{scope}[every node/.append style={xslant=0, yslant=0}]
     \node[above] at (1.2,0.1) {$\mathbb{R}^{d-2}$};
               \end{scope}
\end{tikzpicture}%
\caption{ \label{fig:dim_red}
Dimensional reduction. }
\end{figure}

Dimensional reduction is the statement that, when all insertion points
$y$ appearing in \eqref{eq:expVal} are restricted to the subspace
$\mathbb{R}^{d-2}$, as illustrated in Fig.~\ref{fig:dim_red}, the
correlator can be equivalently computed in a $(d-2)$-dimensional theory.
This statement can be written as
\begin{equation}
\label{dim_red_e.g.}
\boxed{
\phantom{\Bigg|}
\langle \Phi(y) \ldots \rangle_{d|2}
\;\big|_{d-2}
=
\langle \phi(x_{\parallel}) \ldots \rangle_{d-2}
\phantom{\Bigg|}
}\, ,
\end{equation}
where the parallel coordinates $x^{i}_{\parallel}$,
with $i = 1, \ldots, d-2$, parametrize the $(d-2)$-dimensional space, and the projection acts on all the insertion points as follows:
\begin{equation}
\label{def:x,y}
y\big|_{d-2}
\equiv
\bigl( x_{\parallel}, y_{\perp} = 0 \bigr) \, ,
\qquad
y_{\perp} \equiv (\x, \y, \theta, \bar{\theta}) \, ,
\end{equation}
with $\x \equiv x^{d-1}$ and $\y \equiv x^{d}$.
The $(d-2)$-dimensional correlator on the right-hand side of
\eqref{dim_red_e.g.} is defined through the reduced path integral
\begin{equation}
\langle O(x_{\parallel}) \ldots \rangle_{d-2}
=
\frac{1}{\Zred}
\int \mathcal{D}\phi \,
O(x_{\parallel}) \ldots \,
e^{-\Sred[\phi]} \, ,
\end{equation}
with the reduced action given by
\begin{equation}
\label{S_red}
\Sred[\phi]
=
2\pi
\int d^{d-2}x_{\parallel}
\left[
-\frac{1}{2}\,
\phi\,\partial^{2}_{\parallel}\phi
+
V(\phi)
\right] .
\end{equation}

Equation~\eqref{dim_red_e.g.} explicitly states dimensional reduction for
correlators of the fundamental fields $\Phi$ and $\phi$. This statement
can be generalized to composite operators, including operators dressed
with derivatives, as we discuss in the following sections. In the case
of spinning operators, the restriction
$\big|_{{d-2}}$ in \eqref{dim_red_e.g.} also entails a
projection of tensor indices onto $\mathbb{R}^{d-2}$. More precisely, one implements this projection by considering the
branching
$OSp(d|2) \to SO(d-2) \times OSp(2|2)$ and keeping only the
$OSp(2|2)$-singlet components. For instance, a vector $V^{a}$ of
$OSp(d|2)$ decomposes into an $SO(d-2)$ vector for
$a = 1, \ldots, d-2$, and  an $OSp(2|2)$ vector for
$a = d-1, d, \theta, \bar{\theta}$. From here, only the $SO(d-2)$ vector component survives the projection, since it's the only component invariant under $OSp(2|2)$.

Establishing \eqref{dim_red_e.g.} and its generalizations is the main goal of this paper. In the following sections, we will address this problem using supersymmetric tools such as localization and cohomological arguments.

\subsection{Localization: a general review}
\label{sec:loc_review}
Supersymmetric localization is a powerful method that enables exact, nonperturbative evaluations of path integrals in supersymmetric theories. 
The first step in localization is to consider a  fermionic supercharge $Q$. This quantity is often chosen to be nilpotent, satisfying $Q^2=0$; in what follows, however, we relax this condition.
{\Blue It is standard to call operators in the kernel of $Q$ as $Q$-closed, while the ones in the image of $Q$ as  $Q$-exact,  namely 
\begin{equation}
 Q\mbox{-closed}=\{\Ocal:   Q\mathcal{O}=0 \} \, , \qquad
  Q\mbox{-exact}=\{\Ocal: \exists  \mathcal{O'} \textrm{s.t.} \mathcal{O}=Q \mathcal{O'}\} \, .
\end{equation}
}
The procedure for localizing the path integral consists of deforming the action by a $Q$-exact term $S_{\text{loc}} \equiv Q\mathcal{V}$ multiplied by a generic parameter $t$:
\begin{equation}
    Z_t \equiv \int_{\mathcal{F}} \mathcal{D}\Phi e^{-S-tQ \mathcal{V}} \, .
    \label{Zt}
\end{equation}
If the charge $Q$ is not nilpotent, as in our case, it is also necessary to ensure that $Q \mathcal{V}$ is closed:
$Q^2\mathcal{V}=0$. Then it is easy to check that the path integral is invariant under such a deformation:
\begin{equation}
\begin{aligned}
    \frac{d}{dt} Z_t
    &= 
    -\int_{\mathcal{F}} \mathcal{D}\Phi (Q \mathcal{V})e^{-S-tQ \mathcal{V}}
    = -\int_{\mathcal{F}} \mathcal{D}\Phi Q (\mathcal{V}e^{-S-tQ \mathcal{V}}) = 0 \, ,
    \label{Stokes' theorem}
\end{aligned}
\end{equation}
where we used the fact that the action is invariant under the supersymmetry, i.e., $QS=0$. In addition, we assumed that the
integration measure is $Q$-invariant and there are no boundary terms, so that the integral of a total derivative in field space vanishes.

It is further convenient to choose a deformation $Q\mathcal{V}$ whose bosonic part is positive semi-definite. In this way, we can evaluate the partition function by taking the limit $t\to \infty$:
\begin{equation}
   \lim_{\substack{t \to \infty}} Z_t = \lim_{\substack{t \to \infty}}\int_{\mathcal{F}} \mathcal{D}\Phi e^{-S-tQ \mathcal{V}} \, ,
    \label{deformed path integral}
\end{equation}
where the integrand is dominated by the saddle points $\mathcal{F}_S$ of $Q\mathcal{V}$. Crucially, since $ Z_0 =\lim_{\substack{t \to \infty}} Z_t $, the saddle point ``approximation'' here is exact.

To evaluate the path integral (\ref{deformed path integral}), one expands the fields $\Phi$ around the localization locus field configurations $\Phi_S \in \mathcal{F}_S$:
\begin{equation}
\label{eq:Phi=PhiS+deltaPhi}
    \Phi = \Phi_S + \frac{1}{\sqrt{t}} \delta\Phi\, , 
\end{equation}
where $\delta\Phi$ parametrizes directions transverse to the localization locus. In the limit $t \to \infty$, the inverse of $t$ behaves like an effective Planck constant $\hbar_{\text{eff}} = 1/t$, and the semiclassical loop expansion in $\hbar_{\text{eff}}$ reduces to 
\begin{equation}
    S[\Phi]=S[\Phi_S]+S_{\text{loc}}^{(2)}[\delta\Phi] \, ,
\end{equation}
where $S_{\text{loc}}^{(2)} = \frac{1}{2} \int S^{\prime\prime}_{\text{loc}}[\Phi]|_{\Phi=\Phi_S}\delta\Phi^2$ is the quadratic action for fluctuations around the localization locus.\footnote{In practice, since we will choose localization terms $S_{\text{loc}}$ quadratic in the field $\Phi$, the form of $S_{\text{loc}}^{(2)}$ will always coincide with $S_{\text{loc}}$.}
 This is a one-loop exact result since the higher terms are weighted by powers of $1/t$ and vanish in the $t\to\infty$ limit. Thus, the partition function reduces to (and is exactly equal to)
\begin{equation}
    Z = \int_{\mathcal{F}_S} \mathcal{D}\Phi_S e^{-S[\Phi_S]}Z_{\text{1-loop}}[\Phi_S]\,,
    \label{Z_localized}
\end{equation}
where $Z_{\text{1-loop}}$ is the path integral of $\text{exp}({-S_{\text{loc}}^{(2)}})$, which is Gaussian since it is quadratic in the fields $\delta\Phi$. This integral yields a determinant as the result.

The result of the localized path integral~\eqref{Z_localized} does not depend on the choice of deformation term~$S_{\text{loc}}$. We will use this flexibility and pick the term that makes the calculations easiest.

The localization argument can be simply generalized to correlation functions as long as the operator insertions are $Q$-closed. 
This can be seen, for example, by deforming the action by a source term $S_J= S+\int  d^{d}x\,J \Ocal $ (where $J$ is the source) and noticing that as long as $S_J$ is $Q$-closed, one can repeat the same steps above by simply substituting $S \to S_J$. 
We therefore find that all correlators of $Q$-closed operators can be localized. 
Moreover it is easy to see that, within correlators of $Q$-closed insertions, all observables that are both $Q$-exact and $Q$-closed vanish, namely 
\begin{equation}
    \langle {Q(\dots)} \Ocal \rangle=\frac{1}{Z} \int_{\mathcal{F}} \mathcal{D} \Phi Q(\dots) \Ocal e^{-S[\Phi]}=\frac{1}{Z} \int_{\mathcal{F}} \mathcal{D} \Phi Q(\dots \Ocal  e^{-S[\Phi]})=0 \, .
\end{equation}
As a result, correlators of $Q$-closed observables $\Ocal_i$ only depend on their class in equivariant cohomology, namely on the operators modulo $Q$-exact contributions that preserve $Q$-closedness.  Explicitly, for operators $\Ocal'_i$ such that $Q^2\Ocal'_i=0$,
\begin{equation}
\langle\Ocal_1 \dots \Ocal_n \rangle=\langle (\Ocal_1 +Q \mathcal{O}'_1) \dots  (\Ocal_n +Q \mathcal{O}'_n)\rangle \, .
\end{equation}
This defines correlators of operators in the equivariant cohomology of~$Q$.

\subsection{Unleashing \(Q\) in PS models }
\label{sec:unleashQ}

Above, we introduced the idea of localization and (equivariant) cohomology, but we were agnostic about the specific supersymmetry and the explicit form of $Q$.
In the rest of the paper, we will focus on a specific fermionic charge $Q$ given by the following combination of two PS superrotations:\footnote{\Blue The PS supersymmetry is reminiscent of BRST symmetry in gauge-fixed theories \cite{Becchi:1975nq, Becchi:1976nq, Tyutin:1975qk, Bonora:1980pt}, in that both involve spinless fermionic degrees of freedom and naturally lead to a cohomological organization of operators. In the BRST case, this cohomological structure is usually associated with the charge generating translations along Grassmann directions in superspace. In contrast, the charge $Q$ used here is not a fermionic translation, but is built from fermionic rotations inside $OSp(d|2)$. This distinction is important: these rotations can have as a fixed locus $\mathbb R^{d-2}$, which is the geometric origin of the dimensional-reduction argument.}
\begin{equation}
\label{def:Q}
    Q \equiv M^{\x\theta} + M^{\y\bar{\theta}} \, ,
\end{equation}
where $\x = x^{d-1}$  and $\y = x^{d}$ following the conventions of \eqref{def:x,y}.
In particular, the action of $Q$ on a scalar superfield is given by the Killing vector $q \equiv m^{x\theta}+m^{y\bar\theta}$:
\begin{equation}
   Q \Phi(y)=  
   q \Phi(y) 
   = (\x \partial_{\bar{\theta}}-\theta \partial_{\x}-\y \partial_\theta-\bar{\theta} \partial_{\y}) \Phi(y) \, .
\end{equation}
We will study the (equivariant) cohomology with respect to this charge, and we will also use it for the localization arguments.
The supercharge $Q$ is not nilpotent, since it squares to
\begin{equation}
    Q^2  = \tfrac{1}{2}\bigl(M^{\theta \theta}+M^{\bar{\theta} \bar{\theta}}\bigr) + M^{\x\y} \, ,
\end{equation}
as detailed in appendix~\ref{sec: App A}. 
For convenience, we also write the action of $Q$ and $Q^2$ on the superfield components, 
\begin{equation}
\begin{aligned}
\begin{array}{llll}

    Q \varphi=\x \psi-\y \bar{\psi} \, , 
     \; 
    & Q \bar{\psi}=\partial_\x \varphi+\x \omega \,,
    & Q \psi=\partial_\y \varphi+\y \omega \,, 
    & Q \omega=\partial_\y \bar{\psi}-\partial_\x \psi \,,
    \\
    Q^2 \varphi=M^{\x\y} \varphi \,, 
        \; 
    & Q^2 \bar{\psi}=\psi + M^{\x \y} \bar{\psi}\,, 
      \; 
    & Q^2 \psi=-\bar{\psi} + M^{\x \y} \psi\,,
       \; 
    & Q^2 \omega=M^{\x\y} \omega\, \,.
\end{array}
\label{QPhi and Q^2Phi}
\end{aligned}
\end{equation}

While $Q^2\neq 0$, we are interested in studying dimensional reduction to $\mathbb{R}^{d-2}$, where  $Q$ is effectively  nilpotent:
\begin{equation}
    Q^2\big|_{d-2} = 0 \, .
\end{equation}
Consequently, we conclude that the dimensionally reduced operators can be naturally classified according to the cohomology of $Q$. 
{\Blue 
Let us briefly comment on the choice of $Q$ in \eqref{def:Q}. First, $Q$ is a fermionic charge, which is essential for cohomological and localization arguments. We choose $Q$ to be a generator of the subgroup $OSp(2|2) \subset OSp(d|2)$ because this  leaves $\mathbb{R}^{d-2}$ fixed, making it the natural starting point for proving dimensional reduction. This choice is not unique: $OSp(2|2)$ contains four fermionic generators. Although parts of the arguments below would go through for other choices of $Q$, the specific choice in \eqref{def:Q} is particularly well suited to the localization argument, see footnote \ref{otherQ}.
Our choice also matches that of \cite{Zaboronsky:1996qn}, where the same generator was used.}

In the following section, we provide a cohomological argument for dimensional reduction. The idea is to show that the cohomology of $Q$ actually defines the operators in the lower-dimensional theory. The $Q$-exact operators will automatically decouple and we shall make contact with a geometric decoupling argument introduced in~\cite{paperI}.
Then, we will provide three versions of localization arguments. Two of them follow the idea above, but with different localization terms~$S_{\text{loc}}$.
In particular, the argument in \autoref{sec:Localization} is a new proof obtained by following the modern conventions of the localization literature~\cite{Pestun:2016zxk}. The argument in \autoref{sec:Zaboronsky} is a review of Zaboronsky's proof in~\cite{Zaboronsky:1996qn}, which uses a non-standard term~$S_{\text{loc}}$. We shall explain that this choice gives rise to some subtleties.

The third argument in \autoref{sec:Cardy} reviews a different logic presented by  Cardy in~\cite{Cardy:1983aa}. The deformed path integral $Z_t$ chosen by Cardy is not computed by taking a saddle point limit for $t \to \infty$, but instead $Z_t$ interpolates between the original action and the reduced one by appropriately tuning the parameter~$t$. In our review, we aim at rephrasing the argument in terms of the cohomology of~$Q$, and we will further explain how to generalize it to account for other models possessing the same supersymmetry.

\section{Cohomological argument}
\label{sec:cohomology}

In \cite{paperI}, a very simple argument for dimensional reduction was
presented. The first step consists in restricting the theory to
$\mathbb{R}^{d-2}$, which is sometimes referred to as introducing a
trivial defect. Under this restriction, the symmetry is broken as
\begin{equation}
ISO(d|2)
\;\longrightarrow\;
ISO(d-2) \times OSp(2|2) \, .
\end{equation}
The trivial defect preserves Poincar\'e symmetry along the defect, as
well as the $OSp(2|2)$ rotations in the directions transverse to it.
From the point of view of the defect theory, the latter are realized as
a global symmetry. As explained in \autoref{sec:dim_red}, dimensional reduction for
spinning representations comes with a projection onto $OSp(2|2)$
singlets. In the following we thus focus only on this set of operators.

In the kinematics with broken symmetry, the metric $g_{d|2}$ on the
superspace $\mathbb{R}^{d|2}$ similarly decomposes into two parts that
are preserved by the broken symmetry group:
\begin{equation}
g_{d|2}
=
g_{d-2}
+
g_{2|2} \, ,
\end{equation}
where $g_{d-2}$ is the ``parallel'' metric on $\mathbb{R}^{d-2}$, while
$g_{2|2}$ is the ``transverse'' metric on $\mathbb{R}^{2|2}$. More
explicitly, $g_{d-2}^{ab} = \delta^{ab}$ for
$a,b = 1, \ldots, d-2$ and vanishes otherwise, while $g_{2|2}^{ab}$
vanishes for $a,b = 1, \ldots, d-2$ and coincides with $g_{d|2}^{ab}$ on
the remaining components. Both metrics are therefore available as
building blocks for operators and correlation functions.

The central claim of \cite{paperI} is that, within the space of
$OSp(2|2)$ singlets,
\begin{equation}
\label{def:decoupling_KRT}
\begin{minipage}{35em}
\begin{itemize}
\item all operators constructed using $g_{2|2}$ decouple;
\item operators constructed solely using $g_{d-2}$ map to the
dimensionally reduced operators.
\end{itemize}
\end{minipage}
\end{equation}
This result follows from a simple but powerful argument. Correlation
functions of $OSp(2|2)$ singlets must be built from $OSp(2|2)$-invariant
building blocks, yet no non-vanishing invariant can be constructed
using the transverse metric $g_{2|2}$. Indeed, correlators only involve
operators inserted at points $x_\parallel \in \mathbb{R}^{d-2}$, for which
$g^{ab}_{2|2} x_{\parallel\, b} = 0$, and moreover the supertrace of $g_{2|2}$
vanishes. As a result, contracting $g_{2|2}$ with any available object
gives zero, and no invariant can be formed.

While this argument is simple and compelling, the statement
\eqref{def:decoupling_KRT} does not, by itself, identify a precise
mathematical structure underlying the set of decoupled operators. In
particular, it does not explain why operators built out of $g_{2|2}$
should be special. Here we show that the statement \eqref{def:decoupling_KRT} admits a
natural and elegant reformulation in terms of the cohomology of the
supercharge $Q$. More precisely, within the space of $OSp(2|2)$
singlets,
\begin{equation}
\label{def:decoupling_here}
\begin{minipage}{35em}
\begin{itemize}
\item $Q$-exact operators decouple;
\item the $Q$-cohomology defines the  dimensionally
reduced operators.
\end{itemize}
\end{minipage}
\end{equation}
This reformulation is particularly appealing, since the decoupling of
$Q$-exact operators is automatic, and the cohomology of a supercharge
provides a well-defined and intrinsic characterization of the relevant
operator space.

\begin{figure}
\centering
\begin{tikzpicture}[
  font=\normalsize,
  >=Stealth,
  box/.style={
    draw=black,
    thick,
    rounded corners=10pt,
    align=center,
    inner sep=6pt,
    minimum height=6mm
  },
  smallbox/.style={
    box,
    fill= \mycolor
  }
]
\node (qclosedtext) {%
  \begin{tabular}{c}
    \textbf{$Q$-closed} \\[-1mm]
  \end{tabular}
};

\node[
  smallbox,
  right=4mm of qclosedtext,
  minimum width=3.0cm
] (qexact) {%
  \begin{tabular}{c}
    \textbf{$Q$-exact} \\[-1mm]
    \footnotesize (built with $g_{2|2}$)
  \end{tabular}
};

\node[
  box,
  fit=(qclosedtext) (qexact),
  inner sep=8pt
] (qclosedbig) {};

\node[
  box,
  below=25mm of qclosedbig.south,
  minimum width=3cm
] (cft)
{operators of $\text{QFT}_{d-2}$};

\draw[thick] (qexact.south) -- ++(0,-14mm) 
node[
] {}
-- node[pos=0.6, below]{\footnotesize decouple} ++(0,0);

\draw[thick, thick, ->] (qclosedbig.south) -- (cft.north);

\coordinate (midY) at ($(qclosedbig.south)!0.5!(cft.north)$);
\node at ($(qclosedbig.center |- midY)$) { 
};

\end{tikzpicture}

\caption{
\label{fig:cohomology}
Diagram for the cohomological argument: among the $OSp(2|2)$-invariant sector of PS QFT$_{d}$, $Q$-exact operators decouple, while the cohomology of $Q$ maps to the operators of the dimensionally reduced theory.}
\end{figure}
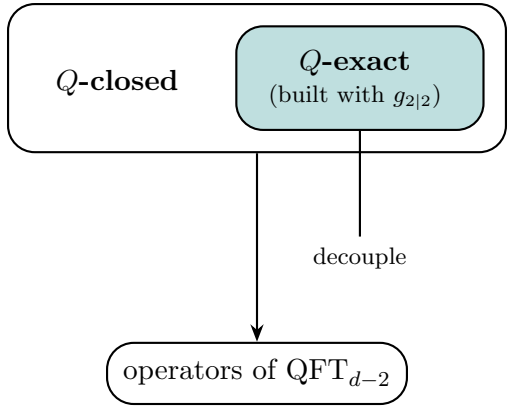

Notice that all $OSp(2|2)$ singlets are automatically $Q$-closed, since
$Q$ is constructed from generators of $OSp(2|2)$. Therefore, in order
to establish the equivalence between \eqref{def:decoupling_KRT} and
\eqref{def:decoupling_here}, it suffices to show that, within the space
of $OSp(2|2)$ singlets,
\begin{equation}
\label{def:proof_cohomology}
\boxed{
\emph{
 \phantom{$\bigg|$} operators built out of  $ g_{2|2}$ \ $\underset{\text{one to one}}{\longleftrightarrow}$ \ $Q$-exact operators 
   \phantom{$\bigg|$}
 }
 }\,.
\end{equation}
In the remainder of this section, we provide an argument in support of
\eqref{def:proof_cohomology}. This establishes that the decoupling
argument of \cite{paperI}, summarized in
\eqref{def:decoupling_KRT}, can be equivalently recast in the
cohomological language of \eqref{def:decoupling_here}, as schematically
illustrated in Fig.~\ref{fig:cohomology}. We will see in \autoref{sec:CohomConseq} that this cohomological viewpoint has several interesting consequences.

\subsection{\(Q\)-exact operators \texorpdfstring{\(\leftrightarrow\)}{<->} operators built  using \texorpdfstring{\(g_{2|2}\)}{g(2|2)}}
To begin, we should explain how the $OSp(2|2)$-invariant operators built out of $g_{2|2}$ look like.   
Some typical examples include
\begin{equation}
    \partial_\perp^2\Phi
    \,
    , \quad \partial_\perp^a\Phi\partial_{\perp a}\Phi
    \,
    ,\quad \partial_\perp^a\Phi^2\partial_{\perp a}\partial^b_\| \Phi
    \,
    ,\quad \partial_\perp^b\partial_\perp^a\Phi\partial_{\perp a} \Phi\partial_{\perp b}\Phi \partial^c_\| \partial^d_\|\Phi
    \, ,
\end{equation}
where the derivative $\partial^{a}_{\parallel }\equiv  g_{d-2}^{ab} \partial_{ b}$ is parallel,  while   
$\partial_{\perp}^a\equiv g_{2|2}^{ba} \partial_{ b} $ is orthogonal to $\mathbb{R}^{d-2}$. These operators are generically built using
any type of parallel derivatives, but crucially, the orthogonal derivatives must always appear contracted to ensure $OSp(2|2)$ invariance. 

A compact way to characterize all $OSp(2|2)$-invariant operators is the
following:
\begin{equation}
\label{def:singlets}
\lim_{y_i \to y}
\bigl(
P_{\parallel}^{\{\mu\}}
P_{\perp}
\,
\Phi(y_1) \cdots \Phi(y_n)
\bigr)\big|_{d-2} \, ,
\end{equation}
where the operators $\Phi(y_i)$ are initially taken at distinct points
$y_i$, and the coincident limit $y_i \to y$ is performed at the end.
A basis of operators is provided by the following differential operators:\footnote{
Strictly speaking, this defines a basis of operators that does not
transform into irreducible representations of the parallel rotation
group. Irreducible representations can be obtained by appropriate
Young symmetrization of the indices and, when necessary, by taking
traces.}
\begin{eqnarray}
P_{\parallel}^{\{\mu\}}
&\equiv&
\partial^{\mu^1_1}_{\parallel 1}
\cdots
\partial^{\mu^1_{\ell_1}}_{\parallel 1}
\cdots
\partial^{\mu^n_1}_{\parallel n}
\cdots
\partial^{\mu^n_{\ell_n}}_{\parallel n} \, ,
\\
P_{\perp}
&\equiv&
\prod_{1 \leq i \leq j \leq n}
\bigl(
\partial_{\perp i} \cdot \partial_{\perp j}
\bigr)^{m_{ij}} \, ,
\label{eq:Pperp}
\end{eqnarray}
where $\partial^{a}_{\parallel i}$ and $\partial^{a}_{\perp i}$ denote
parallel and orthogonal derivatives acting on the point $y_i$,
respectively. To be more explicit about the orthogonal operator, note that
\begin{equation}
\partial_{\perp i} \cdot \partial_{\perp j}
=
\partial_{\x_i}\partial_{\x_j}
+
\partial_{\y_i}\partial_{\y_j}
+
\partial_{\bar{\theta}_i}\partial_{\theta_j}
-
\partial_{\theta_i}\partial_{\bar{\theta}_j} \, .
\end{equation}

The key object in this construction is the differential operator
$P_{\perp}$, which is built from contracted orthogonal derivatives in order to produce an $OSp(2|2)$ singlet. The claim of
\eqref{def:decoupling_KRT} is that whenever $P_{\perp}$ is non-trivial,
namely when at least one $m_{ij} \neq 0$, the corresponding operator
decouples. In what follows, we will show that this decoupling can be
understood cohomologically: when $P_{\perp}$ is non-trivial, the
operator is in fact $Q$-exact.

The operator $P_{\parallel}^{\{\mu\}}$ plays no essential role in this
argument, as it commutes with $Q$ and can therefore be factored out of
all computations. It is thus sufficient to establish the relation
\begin{equation}
\bigl(
P_{\perp}
\,
\Phi(y_1) \cdots \Phi(y_n)
\bigr)\big|_{d-2}
=
Q[\ldots] \, ,
\end{equation}
which would immediately imply that
$
P_{\parallel}^{\{\mu\}}
P_{\perp}
\,
\Phi(y_1) \cdots \Phi(y_n)
\big|_{d-2}$
is $Q$-exact.

We can, in fact, write an explicit expression for the $Q$-exact
operator. To this end, let us consider a generic function
$F \equiv F(y_1, \ldots, y_n)$, which arises from the combination of superfields:
\begin{equation}
F(y_1, \ldots, y_n)
=
P'_{\perp}\,
\Phi(y_1) \cdots \Phi(y_n)  \,,
\end{equation}
where $P'_{\perp}$ is obtained from $P_{\perp}$ by removing a single
factor $(\partial_{\perp i} \cdot \partial_{\perp j})$, that is,
$(\partial_{\perp i} \cdot \partial_{\perp j})\, P'_{\perp}
\equiv
P_{\perp}$.
With this notation, one finds
\begin{eqnarray}
    \boxed{
    (\partial_{\perp i} \cdot \partial_{\perp j})\, F
    \big|_{d-2}
    =
    Q \Bigl[
    (\partial_{\theta_i} \partial_{\x_j} + \partial_{\bar{\theta}_j} \partial_{\y_i} ) F
    \bigr|_{d-2}
    \Bigr]
    } \, ,
\label{eq:QexactBeast}
\end{eqnarray}
as proven in appendix~\ref{App:proof_Q_exact}.
Taking the coincident limit $y_i \to y$, we conclude that all operators
containing contractions of the orthogonal metric $g_{2|2}$ are
$Q$-exact.\footnote{Different choices of which factor to omit in
$P'_{\perp}$ lead to different operators whose $Q$-variation reproduces
the desired expression upon reduction to $\mathbb{R}^{d-2}$.}
As a simple example, consider the operator
$\Phi(y_1)\,\partial^{2}_{\perp y_2}\Phi(y_2)$, for which one finds
\begin{equation}
    \bigl(
    \Phi(y_1)\,\partial^{2}_{\perp y_2}\Phi(y_2)
    \bigr)\big|_{d-2}
    =
    Q\bigl((
    \varphi_1\,\partial_{\y_2}\psi_2
    +
    \varphi_1\,\partial_{\x_2}\bar{\psi}_2
    )|_{d-2}\bigr) \,.
\end{equation}
Another example is provided by the operator
$(\partial_{\perp y_1} \cdot \partial_{\perp y_2})\,
\Phi(y_1)\Phi(y_2)$, which satisfies
\begin{equation}
    \bigl(
    (\partial_{\perp y_1} \cdot \partial_{\perp y_2})\,
    \Phi(y_1)\Phi(y_2)
    \bigr)\big|_{d-2}
    =
    Q\bigl(
    (\psi_1\,\partial_{\y_2}\varphi_2
    +
    \bar{\psi}_1\,\partial_{\x_2}\varphi_2
    )|_{d-2}\bigr) \,.
\end{equation}
We refer the reader to appendix~\ref{App:proof_Q_exact} for further
details and examples.

\subsection{Cohomology and consequences of the decoupling}
\label{sec:CohomConseq}

Above, we showed that all $OSp(2|2)$-singlet operators can be written in
the form~\eqref{def:singlets}, and that any operator with a non-trivial
$P_{\perp}$ is $Q$-exact. It is well known that the correlator of a
$Q$-exact operator with any number of $Q$-closed operators vanishes:
\begin{equation}
\langle
\widetilde{\mathcal{O}}
\,
\mathcal{O}_1 \cdots \mathcal{O}_n
\rangle
=
0 \, ,
\end{equation}
where all operators are $OSp(2|2)$ singlets, the operators
$\mathcal{O}_i$ are $Q$-closed, and $\widetilde{\mathcal{O}}$ is
$Q$-exact. The cohomological argument, therefore, explains why all
operators built out of the orthogonal metric $g_{2|2}$ decouple, in
agreement with the results of~\cite{paperI}.

We can thus characterize the equivalence classes of operators in
$Q$-cohomology by restricting to the operators~\eqref{def:singlets} with
trivial $P_{\perp}$. These operators are in one-to-one correspondence
with the operators of the dimensionally reduced theory, which are
spanned by
\begin{equation}
\lim_{x_i \to x}
P_{\parallel}^{\{\mu\}}
\,
\varphi(x_{1\parallel}) \cdots \varphi(x_{n\parallel}) \, .
\end{equation}
This establishes that the $Q$-cohomology precisely reproduces the set of
operators of the reduced theory.

Let us now turn to the stress tensor. Rephrasing the argument
of~\cite{paperI}, we show that the existence of a conserved superstress
tensor implies, in cohomology, the existence of the stress tensor of the
dimensionally reduced theory. Consider a superstress tensor
$\mathcal{T}_{ab}$ satisfying the conservation equation
$\partial^{a} \mathcal{T}_{ab} = 0$. Splitting the indices into parallel
and orthogonal components, one finds
\begin{equation}
\partial^{a} \mathcal{T}_{a\nu}(x_{\parallel})
=
g_{d-2}^{ij}\,
\partial_{i} \mathcal{T}_{j\nu}(x_{\parallel})
+
g_{2|2}^{ab}\,
\partial_{a} \mathcal{T}_{b\nu}(x_{\parallel}) \, .
\label{eq:TCons}
\end{equation}
Here, the second term is $Q$-exact by the arguments presented above, and therefore
\begin{equation}
0
=
\partial^{a} \mathcal{T}_{a\nu}(x_{\parallel})
=
\partial^{i} \mathcal{T}_{i\nu}(x_{\parallel})
+
Q[\ldots] \, ,
\end{equation}
with $i$ running over the parallel directions. Since the operator $\mathcal{T}$ is inserted at $x_{\parallel}$, only
the lowest component of the superfield contributes, which has scaling
dimension $d-2$ \cite{paperI}. This implies the existence, in $Q$-cohomology, of a
spin-two operator of dimension $d-2$ that is conserved. This operator
can be identified with the stress tensor of the dimensionally reduced
theory. Similar remarks apply to other conserved currents.

In the case of CFTs, the
superstress tensor is required to be supertraceless. Using
\begin{equation}
\mathcal{T}_{ab} g_{2|2}^{ab}
+
\mathcal{T}_{ij} g_{d-2}^{ij}
=
0 \, ,
\end{equation}
and noting that the $2|2$ contribution decouples in cohomology, it
follows that the reduced stress tensor is traceless. We thus conclude
that the $Q$-cohomology contains the conserved and traceless stress
tensor of the dimensionally reduced theory.\footnote{This construction
can be checked explicitly in free theory using the superstress tensor
\begin{equation}
\mathcal{T}_{ab}
=
\left(
\partial_{(a} \Phi
\right)
\left(
\partial_{b]} \Phi
\right)
-
\frac{d-4}{d-2}\,
\Phi\,\partial_{a}\partial_{b}\Phi
-
\frac{g_{ba}}{d-2}\,
\left(
\partial^{c}\Phi
\right)
\left(
\partial_{c}\Phi
\right) ,
\end{equation}
where $(ab]$ denotes graded symmetrization.} Furthermore, one can also adapt the argument
of~\cite{paperI} to show that the operator
product expansion of $OSp(2|2)$ singlets descends consistently to the
reduced OPE in $Q$-cohomology.

The use of supercharge cohomology to define effective theories on
lower-dimensional manifolds is not new. In four-dimensional
$\mathcal{N}=2$ superconformal field theories, it was shown in
\cite{Beem:2013sza} that the cohomology of a particular supercharge can
be used to define a two-dimensional chiral CFT. This
construction is closely reminiscent of Parisi--Sourlas dimensional
reduction.\footnote{While reminiscent, the two setups differ in
important ways. In the case of \cite{Beem:2013sza}, the
higher-dimensional theory is unitary, whereas the lower-dimensional one
is not, which is precisely the opposite of the situation considered
here. Moreover, our construction applies in arbitrary dimensions, and
the resulting dimensionally reduced theory can be significantly richer,
as its operator content is not limited to conserved currents, in
contrast to chiral theories.}
Similar cohomological reductions have subsequently been identified in
theories in various dimensions and with different amounts of
supersymmetry, see e.g.~\cite{Beem:2014kka, Chester:2014mea}. From this
point of view, the reformulation presented here places these apparently
distinct constructions on a common footing, emphasizing the role of
supercharge cohomology as a unifying principle.

\section{Localization argument}
\label{sec:Localization}

In this section, we present a standard localization argument for
dimensional reduction in PS supersymmetric QFTs, which, to the best of
our knowledge, has not been discussed explicitly in the existing
literature. For simplicity, we focus on the standard scalar PS action \eqref{eq:PS_action}.
Nevertheless,  similar arguments are expected to extend to more generic QFTs possessing
PS supersymmetry.\footnote{In this respect, Cardy’s interpolation
argument, discussed in the next section, is more general, as it does not
rely on the explicit form of the action.}

Following the general localization logic reviewed in
\autoref{sec:loc_review}, we deform the path integral by a $Q$-exact
term, $S_{\text{loc}} = Q \mathcal{V} $.
The deformed partition function then reads
\begin{equation}
Z_{\text{SUSY},t}
=
\int
\mathcal{D}\varphi \,
\mathcal{D}\omega \,
[\mathcal{D}\bar{\psi}]\,
[\mathcal{D}\psi]\,\,
e^{
- S_{\text{SUSY}}[\varphi,\bar{\psi},\psi,\omega]
- t\, Q\mathcal{V} }  ,
\label{eq: ZSUSY,t}
\end{equation}
where $S_{\text{SUSY}}$ is given in \eqref{eq:PS_action} and $t$ is a real
deformation parameter. The functional $\mathcal{V}$ is chosen as
\begin{equation}
\mathcal{V}
=
\int d^{d}x\, \bigl( \psi\, Q\psi + \bar{\psi}\, Q\bar{\psi} \bigr) \, ,
\label{def:locTerm}
\end{equation}
in close analogy with standard localization constructions in SUSY QFTs,
see e.g.~\cite{Pestun:2016zxk} and references therein.
This choice of $\mathcal{V}$ satisfies the standard requirements for
localization. First, one can verify that $Q^{2}\mathcal{V} = 0$ (see
appendix~\ref{sec: App B}). Second, the bosonic part of the deformation
$Q\mathcal{V}$ is positive definite. Indeed, one finds
\begin{align}
S_{\text{loc}}
&=
\int d^{d}x\,\bigl(
(Q\psi)^{2} + (Q\bar{\psi})^{2}
- \psi\, Q^{2}\psi - \bar{\psi}\, Q^{2}\bar{\psi} \bigr)
\label{S_loc}
\\
&=
\int d^{d}x\,\Bigl(
\underbrace{
( \partial_{\x}\varphi + \x \omega )^{2}
+ ( \partial_{\y}\varphi + \y \omega )^{2}
}_{\text{bosonic part } \geq 0}
+
\underbrace{
\psi(\bar{\psi} - m^{\x\y}\psi)
- \bar{\psi}(\psi + m^{\x\y}\bar{\psi})
}_{\text{fermionic part}}
\Bigr) \, ,
\label{S_loc_explicit}
\end{align}
where the second line follows directly from the supersymmetry transformations~\eqref{QPhi and Q^2Phi}.
As discussed in \autoref{sec:loc_review}, these properties ensure that
the deformed partition function \eqref{eq: ZSUSY,t} is independent of
the parameter $t$, and that the limit $t \to \infty$ is well defined.

In the limit $t \to \infty$, the path integral localizes onto the locus
of field configurations for which $S_{\text{loc}}$ vanishes. We denote
this set of saddle-point configurations of $S_{\text{loc}}$ by
$\mathcal{F}_S$. Within the standard SUSY localization framework, one
typically considers
\begin{equation}
\mathcal{F}_S
=
\left\{
\Phi \;\middle|\;
\text{fermions} = 0,
\;
Q(\text{fermions}) = 0
\right\} .
\label{eq:Q-invariant-configurations}
\end{equation}
Requiring $Q(\text{fermions}) = 0$ guarantees that the bosonic part of
$S_{\text{loc}}$ vanishes, as is evident from~\eqref{S_loc}. We also impose the standard
condition $\text{fermions} = 0$, which ensures that the saddle point is
purely bosonic and can be interpreted as a classical field
configuration.\footnote{In the next subsection we will explain that,
strictly speaking, this condition should be relaxed: one should allow
for a non-trivial fermionic profile in order to correctly account for
certain zero modes. We will nevertheless show that this refinement does
not affect the final result, as the corresponding fermionic modes
ultimately decouple. For the moment, we therefore adopt the standard
choice $\text{fermions} = 0$.}

Using the relations \eqref{QPhi and Q^2Phi}, the localization locus
$\mathcal{F}_S$ consists of field configurations
$(\varphi_S,\bar{\psi}_S,\psi_S,\omega_S)$ satisfying
\begin{equation}
\partial_{\x}\varphi_S = -\x \omega_S  \,,
\qquad
\partial_{\y}\varphi_S = -\y \omega_S  \,,
\qquad
\psi_S = 0 \,,
\qquad
\bar{\psi}_S = 0  \,.
\label{localization locus}
\end{equation}
To solve these equations, we multiply the first and second equations by
$\y$ and $\x$, respectively, and subtract them, obtaining
\begin{equation}
\label{eq:Saddle_1}
(\x \partial_{\y} - \y \partial_{\x}) \varphi_S = 0 \,.
\end{equation}
A second relation follows by multiplying the same equations by $\x$ and $\y$, respectively, and adding them:
\begin{equation}
\label{eq:Saddle_2}
(\x \partial_{\x} + \y \partial_{\y}) \varphi_S
= - (\x^{2} + \y^{2}) \omega_S \,.
\end{equation}
Equation \eqref{eq:Saddle_1} implies that $\varphi_S$ is invariant under the
rotations generated by $m^{\x\y}$. We may therefore regard $\varphi_S$
as a function of $x_{\parallel}\in \mathbb{R}^{d-2}$ and the radial coordinate $r=\sqrt{\x^2+\y^2}$ only: 
$\varphi_S = \varphi_S(x_{\parallel}, r, \eta = 0)$. Here, we introduced polar coordinates $(r,\eta)$ in the transverse
plane via $\x = r \cos\eta$, $\y = r \sin\eta$, for which
$m^{\x\y} = \partial_{\eta}$. The second equation then implies
$r \partial_r \varphi_S = - r^{2} \omega_S$.
Altogether, the localization locus is characterized by
\begin{equation}
\varphi_S = \varphi_S(x_{\parallel}, r, \eta = 0)\,,
\qquad
\omega_S = - \frac{1}{r} \partial_r \varphi_S \,,
\qquad
\bar{\psi}_S = \psi_S = 0 \, .
\label{localization locus phi}
\end{equation}

To evaluate the deformed path integral \eqref{eq: ZSUSY,t}, we expand the
fields around the localization locus~\eqref{localization locus phi} by
rescaling the fluctuations as follows
\begin{equation}
\begin{aligned}
\varphi=
\varphi_S + \frac{1}{\sqrt{t}} \delta\varphi 
\, ,
\qquad
\omega=- \frac{1}{r} \partial_r \varphi_S
+ \frac{1}{\sqrt{t}} \delta\omega 
\, ,
\qquad 
\psi
=
\frac{1}{\sqrt{t}} \delta\psi \, 
,
\qquad
\bar{\psi}
=
\frac{1}{\sqrt{t}} \delta\bar{\psi} \, ,
\end{aligned}
\label{saddle point configurations}
\end{equation}
where the fields with the deltas parametrize directions transverse to the
localization locus.
Taking the limit $t \to \infty$, the path integral reduces to
\begin{equation}
Z_{\text{SUSY}}
=
Z_{\text{1-loop}}
\int \mathcal{D}\varphi_S \;
e^{-S[\varphi_S]} \, ,
\label{Z-general}
\end{equation}
where the one-loop determinant, after redefining $\delta\varphi\to\varphi ,\delta\psi\to\psi,\delta\bar{\psi}\to \bar{\psi},\delta\omega\to \omega$,  is given by
\begin{equation}
Z_{\text{1-loop}}
=
\int
\mathcal{D}\varphi\,
\mathcal{D}\omega\,
[\mathcal{D}\bar{\psi}]\,
[\mathcal{D}\psi]\;
e^{-S^{(2)}_{\text{loc}}} \, ,
\label{one loop det}
\end{equation}
where $S^{(2)}_{\text{loc}}$ has the same form as~\eqref{S_loc}. Since
$S^{(2)}_{\text{loc}}$ is a Gaussian action independent of the
 field $\varphi_S$, $Z_{\text{1-loop}}$ is a pure number. Its precise
value will be discussed in the next section, but it is ultimately
unimportant, as it cancels in the computation of  correlation
functions.\footnote{Strictly speaking, $Z_{\text{1-loop}}$ contains a
zero mode and therefore vanishes. In the next section we explain how to
properly treat this zero mode and show that the resulting one-loop
determinant can be consistently taken to be $Z_{\text{1-loop}} = 1$.}

We now compute the localized action $S[\varphi_S]$ appearing in
\eqref{Z-general} and show that it reproduces the reduced action $\Sred$.
Evaluating the PS action \eqref{eq:PS_action} on the localization locus
\eqref{localization locus phi}, we obtain
\begin{equation}
\begin{aligned}
S[\varphi_S]
=
\int d^{d}x \, \bigl(
\partial_{\x}\omega_S \partial_{\x}\varphi_S
+
\partial_{\y}\omega_S \partial_{\y}\varphi_S
+
\partial^{i}\omega_S \partial_{i}\varphi_S
-
\omega_S^{2}
+
\omega_S V^{\prime}(\varphi_S) \bigr) \, ,
\label{reduced action}
\end{aligned}
\end{equation}
where we have separated the kinetic term into contributions along
$\mathbb{R}^{d-2}$ and the transverse directions.

Using the localization equations \eqref{localization locus}, the first two
terms in \eqref{reduced action} can be rewritten as
\begin{equation}
\begin{aligned}
\int d^{d}x\,\bigl(
\partial_{\x}\omega_S \partial_{\x}\varphi_S
+ \partial_{\y}\omega_S \partial_{\y}\varphi_S \bigr)
=
\int d^{d}x\,\bigl(
- \x \frac{1}{2}\partial_{\x}\omega_S^{2}
- \y \frac{1}{2}\partial_{\y}\omega_S^{2} \bigr)
= \int d^{d}x\, \omega_S^{2} \, ,
\end{aligned}
\label{omega_F}
\end{equation}
where in the last step we performed an integration by parts. This precisely
cancels the $-\omega_S^{2}$ term in \eqref{reduced action}. We are therefore
left with
\begin{equation}
\begin{aligned}
S[\varphi_S]
= \int d^{d}x \, \bigl(
\partial^{i}\omega_S \partial_{i}\varphi_S
+ \omega_S V^{\prime}(\varphi_S) \bigr)
= \int d^{d}x\,\Bigl[
-\frac{1}{2r}\partial_r \bigl( \partial^{i}\varphi_S \partial_{i}\varphi_S \bigr)
- \frac{1}{r}\partial_r \varphi_S V^{\prime}(\varphi_S)
\Bigr] .
\end{aligned}
\label{eq:SofBosonSaddle}
\end{equation}

We now decompose the integration over $\mathbb{R}^d$ into
$\mathbb{R}^{d-2}$ and the transverse $\x$-$\y$ plane. Since $\varphi_S$ is independent of $\eta$, the
angular integration is trivial and yields
\begin{equation}
S[\varphi_S]
= 2\pi \int d^{d-2} x_{\|}
\int_0^{\infty} dr \Bigl[
-\frac{1}{2}\partial_r
\bigl( \partial^{i}\varphi_S \partial_{i}\varphi_S \bigr)
- \partial_r V(\varphi_S)
\Bigr] .
\label{eq:SAlmostRedu}
\end{equation}
Assuming that the fields vanish sufficiently fast at infinity and that
$V(0)=0$, the $r$ integral can be performed explicitly, yielding
\begin{equation}
S_{d-2}
= 2\pi \int d^{d-2}x_{\|} \Bigl[
-\frac{1}{2} \phi\,\partial_{\|}^{2}\phi
+ V(\phi) \Bigr] ,
\label{eq:SredFinal}
\end{equation}
where we defined $\phi \equiv \varphi_S(x_{\|},0,0)$. This action coincides
precisely with the reduced action \eqref{S_red}, thus completing the proof of
dimensional reduction.\footnote{The structure here is the same as for the reduction of 3d $\mathcal{N}=4$ superconformal theories to a 1d topological subsector: as observed in~\cite{Dedushenko:2016jxl}, supersymmetric localization eliminates an angular dependence of the 3d fields, and the resulting 2d action is a total derivative that reduces to a boundary term, which serves as the action for the 1d subsector.}\,%
\footnote{\label{otherQ}\Blue Notice that the specific choice of $Q$ is essential for this result. For instance, if one chooses $Q=M^{\x\theta}$, the localization locus is modified: instead of \eqref{localization locus}, one obtains
$\psi=\psib=0$ and $\partial_\x\varphi+\x \omega=0$.
The saddle-point action can then be computed explicitly, and one immediately sees that the preceding argument no longer applies. Indeed, $\partial_\y\varphi$ remains unconstrained. Already in the first step, \eqref{omega_F}, one cannot replace $\partial_\y \varphi$, and hence cannot obtain the right-hand side. Moreover, one can no longer conclude that $\varphi$ depends only on the radial direction, as in \eqref{localization locus phi}, which is a crucial step in the proof of dimensional reduction, since it already implies that the fields are independent of one spatial direction.}

To be precise, the derivation above glosses over a few subtleties that we
address in the next section. In particular, one can check that the one-loop
determinant \eqref{one loop det} contains a zero mode and therefore vanishes,
rendering the computation naively ill-defined. In the next section, we explain
how to correctly account for this zero mode. We will show that it does not
modify the final result \eqref{eq:SredFinal}, and that the properly defined
one-loop determinant is equal to one.

\subsection{1-loop determinant and fermionic zero modes}
\label{sec:1_loop}

In the discussion above, we focused on saddle-point configurations with
vanishing fermionic profiles. This treatment is, however, slightly too
naive due to the presence of fermionic zero modes in the localization
term.\footnote{In the recent computation of the elliptic genus of two-dimensional
$\mathcal{N}=(0,1)$ theories~\cite{Bao:2025xhl}, similar fermionic zero modes of the deformation term were found, which are lifted only by the classical action.}
Indeed, the fermionic part of \eqref{S_loc_explicit} also vanishes
whenever the fermions satisfy
\begin{equation}
\label{cond_zero_modes}
\bar{\psi} = m^{\x\y}\psi \, ,
\qquad
\psi = - m^{\x\y}\bar{\psi} \, .
\end{equation}
As a consequence, even in the limit $t \to \infty$, the fermionic modes
of $S_{\text{SUSY}}$ obeying \eqref{cond_zero_modes} are not suppressed
and must be included in the saddle-point analysis.

To account for these zero modes, we first determine the fermionic
profiles $\psi_S$ and $\bar{\psi}_S$ satisfying
\eqref{cond_zero_modes}. They take the form
\begin{equation}
\label{eq:psiS}
\psi_S(x)
=
\sin\eta\, \bar{\xi}(x_{\|},r)
+
\cos\eta\, \xi(x_{\|},r) \, ,
\quad
\bar{\psi}_S(x)
=
-\sin\eta\, \xi(x_{\|},r)
+
\cos\eta\, \bar{\xi}(x_{\|},r) \, ,
\end{equation}
where $\xi$ and $\bar{\xi}$ are fermionic fields depending on
$x_{\|}$ and $r$, and not the angle~$\eta$.

We then repeat the saddle-point computation using the same bosonic
configurations $\varphi_S$ and $\omega_S$ as in
\eqref{localization locus phi}, but now including the non-trivial
fermionic profile \eqref{eq:psiS}. Substituting these expressions into
the action $S_{\text{SUSY}}$ defined in \eqref{eq:PS_action}, we obtain
the saddle-point action
\begin{equation}
\label{final_reduced_action}
S[\phi,\xi,\bar{\xi}]
=
\Sred[\phi]
+
S_{d-1}[\xi,\bar{\xi},\varphi_S] \, ,
\end{equation}
where $\Sred$ is the reduced action \eqref{eq:SredFinal} obtained in the
previous section by radial integration, and
\begin{align}
\label{eq:S_d-1}
S_{d-1}[\xi,\bar{\xi},\varphi_S]
\equiv
2\pi
\int d^{d-2}x_{\|}
\int_{0}^{\infty} dr\,
\Bigl(
\partial^{i}\xi\, \partial_{i}\bar{\xi}
+
\partial_{r}\xi\, \partial_{r}\bar{\xi}
+
\frac{\xi\bar{\xi}}{r^{2}}
+
\xi\bar{\xi}\, V''(\varphi_S)
\Bigr)
\end{align}
is the additional contribution arising from the fermionic zero modes.
A detailed derivation is provided in appendix~\ref{App:Fermionic_Saddle}.

At first sight, the presence of $S_{d-1}[\xi,\bar{\xi},\varphi_S]$ appears
to obstruct dimensional reduction. However, we
show that $S_{d-1}$ is in fact $Q$-exact and hence decouples from all
physical observables.
To this end, we evaluate the SUSY variation
$Q\varphi_S = \x\,\psi_S - \y\,\bar{\psi}_S$, that is, the transformation
rules \eqref{QPhi and Q^2Phi} applied to the saddle-point configurations
\eqref{localization locus phi} and \eqref{eq:psiS}. This yields the simple
relation
\begin{equation}
\begin{aligned}
Q\varphi_S
= r\,\xi \, .
\end{aligned}
\end{equation}
It follows immediately that $\xi = Q\!\left(\frac{\varphi_S}{r}\right)$, and is therefore $Q$-exact. One can similarly check that $Q\xi = Q\bar{\xi} = 0$.
Using these relations, the additional fermionic contribution to the
saddle-point action can be rewritten as (see appendix \ref{App:Fermionic_Saddle})
\begin{equation}
\!\! S_{d-1}
=
Q\!\left[
2\pi
\int d^{d-2}x_{\|}
\int_{0}^{\infty} dr \,
\left(
\partial^{i}\!\left(\frac{\varphi_S}{r}\right)\partial_{i}\bar{\xi}
+
\partial_{r}\!\left(\frac{\varphi_S}{r}\right)\partial_{r}\bar{\xi}
+
\frac{\varphi_S\,\bar{\xi}}{r^{3}}
+
\frac{V'(\varphi_S)\,\bar{\xi}}{r}
\right)
\right] ,
\end{equation}
which makes it manifest that $S_{d-1}$ is $Q$-exact. As a result, this
term decouples from all physical observables and does not affect the
derivation of the saddle-point action presented in the previous section.
We therefore conclude that, after properly accounting for fermionic zero
modes, the localized action indeed coincides with the reduced action.

Let us now turn to the computation of the one-loop determinant defined in
\eqref{one loop det}. Using radial coordinates and suitable field
redefinitions, the quadratic localization action can be written as
$S^{(2)}_{\text{loc}} = S_b + S_f$, with (see
appendix~\ref{sec:AppC})
\begin{align}
S_b
&=
\int d^{d-2}x_\|\, r\,dr\,d\eta\;
\bigl(
\omega^2
-
\varphi\,\partial_\eta^2 \varphi
\bigr) \, ,
\\
S_f
&=
\int d^{d-2}x_\|\, r\,dr\,d\eta\;
\bigl(
\psi\,\partial_\eta \psi
-
2 \psi\,\bar{\psi}
+
\bar{\psi}\,\partial_\eta \bar{\psi}
\bigr) \, .
\label{eq:SbandSf}
\end{align}

To evaluate the corresponding path integral, we expand each field in
Fourier modes along the $\eta$ direction:
\begin{equation}
\begin{aligned}
\varphi(x_\|,r,\eta)
&=
\sum_{\substack{k=-\infty\\k\neq 0}}^{\infty}
\varphi_k(x_\|,r)\, e^{ik\eta}  \,,
\qquad
\omega(x_\|,r,\eta)
=
\sum_{k=-\infty}^{\infty}
\omega_k(x_\|,r)\, e^{ik\eta} \,,
\\
\psi(x_\|,r,\eta)
&=
\sum_{k=-\infty}^{\infty}
\psi_k(x_\|,r)\, e^{ik\eta}  \,,
\qquad
\bar{\psi}(x_\|,r,\eta)
=
\sum_{k=-\infty}^{\infty}
\bar{\psi}_k(x_\|,r)\, e^{ik\eta}  \,.
\end{aligned}
\label{eq:fermFourier}
\end{equation}

The path integral in \eqref{one loop det} computes fluctuations orthogonal to the saddle-point configurations.
In particular, the $\eta$-independent bosonic modes $\varphi_0$ are already accounted for by $\varphi_S$ and $\omega_S$.
Two of the four fermionic modes with $k=\pm 1$ must be eliminated. 
or this purpose, we change variables from 
$(\psi_{-1},\psi_1,\bar\psi_{-1},\bar\psi_1)$ to $(\alpha^{+}_{-1},\alpha^{+}_1,\alpha^{-}_{-1},\alpha^{-}_1)$ where
$\alpha^{\pm}_{\sigma} \equiv (\bar\psi_{\sigma}\mp i \sigma\psi_{\sigma})/\sqrt{2} $. This map has a trivial Jacobian, so that the path integration measure is unchanged.
In these variables, the zero-mode profiles~\eqref{eq:psiS} take the form $\alpha^{+}_{-1}=\alpha^{+}_1=0$.  
Substituting the Fourier expansions \eqref{eq:fermFourier} into the
action and performing the $\eta$ integral, we obtain
\begin{equation}
\label{Sb_Sf_fourire}
\begin{aligned}
S_b
&= 
2\pi
\int d^{d-2}x_\|\, r\,dr
\sum_{k \neq 0}
\bigl(
\omega_{-k}\omega_k
+
\varphi_{-k} k^2 \varphi_k
\bigr) \, ,
\\
S_f
&= 
2\pi\!
\int d^{d-2}x_\|\, r \, dr \biggl(
4 \alpha^{+}_{-1} \alpha^{+}_1
+
\sum_{k \neq \pm 1} \bigl(
\psi_{-k} i k \psi_k
+ \bar{\psi}_{-k} i k \bar{\psi}_k
- 2 \psi_{-k}\bar{\psi}_k
\bigr) \biggr) .
\end{aligned}
\end{equation}

Both the bosonic and fermionic sectors are quadratic by construction, so the
corresponding path integrals are Gaussian. For each fixed $(x_\|,r)$,
the bosonic integral produces a determinant, while the fermionic one
gives a Pfaffian. Formally multiplying the contributions from all
$(x_\|,r)$ modes,\footnote{Strictly speaking, the product over
$(x_\|,r)$ runs over continuous variables, so
\eqref{Z_1loop=1} should be understood as the continuum limit of an
appropriate discretization. We gloss over these technicalities here,
since the contribution from each mode is equal to one.} we obtain (see
appendix~\ref{sec:AppC} for details)
\begin{equation}
\label{Z_1loop=1}
    Z_{\text{1-loop}}
    =
    \prod_{x_\|,r}
    \biggl(
    \frac{
    2 \prod_{k\geq 0,\; k \neq 1}
    \lvert k^2 - 1 \rvert
    }{
    \prod_{k\geq 1}
    k^2
    }
    \biggr)
    =
    1 \, .
\end{equation}
Here, the numerator arises from the fermionic Pfaffian, while the denominator comes from the bosonic determinant. Writing
$\lvert k^2 - 1 \rvert = \lvert k-1 \rvert (k+1)$, one sees that the
factors cancel pairwise between the numerator and the denominator, yielding a
unit result. It is crucial in this computation to treat separately the
fermionic zero modes with $|k|=1$, as including them would lead to a
vanishing one-loop determinant.

With these subtleties under control, we conclude that the localization
procedure developed in this section provides a complete and consistent
proof of dimensional reduction.

\subsection{Review of Zaboronsky localization}
\label{sec:Zaboronsky}
We now turn to the localization approach introduced by Zaboronsky
\cite{Zaboronsky:1996qn}. This construction involves a simpler
deformation term than~\eqref{S_loc}, but also comes with some conceptual
subtleties, which we discuss below.

The starting point is to project $\Phi=\varphi+\theta\bar\psi+\bar\theta\psi+\theta\bar\theta\omega$ to its restriction $\phi=\varphi|_{\mathbb{R}^{d-2}}$, then extend this lower-dimensional field to a superfield $
\Phi_S
$ on $\mathbb{R}^{d|2}$ that is $Q$-invariant.
For definiteness, we choose
\begin{equation}
\Phi_S(x,\theta,\bar{\theta}) = \phi(x_{\parallel}) \,,
\end{equation}
whose only non-vanishing superfield component is the lowest component, which furthermore is constant in $(r,\eta)$.
Zaboronsky’s deformation term can then be understood as a Gaussian
superfield action centered at $\Phi_S$, namely
\begin{equation}
    S_{\text{loc}}= \int d^d x d \bar{\theta} d \theta (\Phi-\Phi_S)^2
    = \int  d^d x \Bigl[(\varphi-\phi) \omega+\psi \bar{\psi}\Bigr] \, ,
\end{equation}
which is much simpler than~\eqref{S_loc}. 
This choice satisfies $Q S_{\text{loc}}=0$.
In fact, the deformation term fits more directly in the standard localization approach by noting that it is $Q$-exact up to a boundary term that vanishes thanks to $\Phi=
\Phi_S
$ at $r=0$:
\begin{equation}
    (\varphi-\phi) \omega+\psi \bar{\psi}
    =
     Q\left(\frac{\varphi-\phi}{r^2}(\x \bar{\psi}+\y \psi)\right)
    -\frac{1}{2r} \partial_r\left(\varphi-\phi\right)^2 .
\end{equation}
Glossing over the fact that this action is not positive-definite, one can follow the localization playbook and consider the saddles of this action.
A straightforward analysis shows that
the corresponding saddle-point configuration is
\begin{equation}
    \varphi= \phi \,, \quad \omega=0 \,, \quad \psi = 0 \,, \quad \bar{\psi}=0 \,.
\end{equation}
When evaluated on this configuration, the PS Lagrangian in
\eqref{eq:PS_action} vanishes identically. However, the action
\eqref{eq:PS_action} still contains integrals over the  $\x$-$\y$ plane which lead to divergences. As a
result, the localization procedure based on this deformation is
subtle, and the corresponding computation is not fully well defined.

There is, however, a way to recover the desired result, as we now
explain. For large values of $t$, the localization term
$e^{-t S_{\text{loc}}}$ behaves as a delta functional on the space of
superfield configurations, effectively localizing the path integral
onto $\Phi_S$. From this perspective, one can replace
$\Phi \to \phi$ in the superfield action
\eqref{eq:PS_action_superfield}, which then reduces to
\begin{equation}
S
=
V_{2|2}
\times
2\pi
\int d^{d-2}x_{\parallel}
\bigl[
(\partial_{\parallel} \phi)^2 + V(\phi)
\bigr] .
\end{equation}
This expression coincides with the reduced action, up to an overall
factor given by the volume $V_{2|2}$ of the transverse superspace
$\mathbb{R}^{2|2}$, defined as
\begin{equation}
V_{2|2}
\equiv
\int [d\bar{\theta}][d\theta] \, d x \, d y \; 1 \, .
\end{equation}
The integral $V_{2|2}$ is not well defined: it vanishes due to the
fermionic integrations, while diverging because of the bosonic ones.
Nevertheless, one may introduce a regularized volume,
\begin{equation}
V_{2|2}^{\zeta}
\equiv
\int [d\bar{\theta}][d\theta] \, d x \, d y \;
e^{-\zeta (x^{2} + y^{2} - 2 \theta \bar{\theta})}
=
1 \, ,
\qquad (\zeta > 0) \,,
\end{equation}
which is finite and equal to one for $\zeta>0$. Although
$V_{2|2}^{\zeta}$ is strictly defined only for $\zeta>0$, one may argue
by analytic continuation that
$V_{2|2}^{\zeta=0} = V_{2|2} = 1$. Under this assumption, the saddle-point
action reproduces the reduced action, as desired.

Let us also briefly comment on the one-loop determinant. This arises
from the evaluation of the partition function associated with the
Gaussian action
\begin{equation}
\label{S_loc_Zab}
S_{\mathrm{loc}}
=
\int d^{d}x\,\bigl(
2 \varphi \, \omega + 2 \psi \bar{\psi} \bigr) \, ,
\end{equation}
where the field $\varphi$ fluctuates only along the directions
orthogonal to the localization locus.

Since the term $\varphi \omega$ is not positive definite, the
corresponding one-loop determinant is divergent and requires
regularization. One possible prescription is to choose the integration
contour for $\omega$ to be purely imaginary, which yields a finite
one-loop determinant equal to one. We do not dwell further on this
point, as the Gaussian integral contributes only an overall numerical
factor, which cancels in normalized correlation functions.

In summary, while Zaboronsky's argument ultimately leads to the correct
result, the proof involves several subtleties. First, the bosonic part
of the deformation term is not positive definite, which obscures the
standard saddle-point analysis and renders the one-loop determinant
ill-defined. Second, it is essential to work directly in superspace,
since the component Lagrangian vanishes on the localization locus.
Third, the final result is proportional to the volume $V_{2|2}$, which
is only defined through analytic continuation. In view of these
subtleties, one may prefer the localization argument presented in
section~\ref{sec:Localization}.

\section{Interpolation argument} \label{sec:Cardy}
In this section, we discuss an alternative approach to dimensional reduction proposed by Cardy \cite{Cardy:1983aa}. The idea is to introduce a deformation term such that the action interpolates between the SUSY action and the reduced action. Using symmetry arguments in superspace, Cardy argues that the path integral is independent of this deformation, thereby proving dimensional reduction. After briefly reviewing his original argument, we show that his deformation term is, in fact, $Q$-exact. This provides a natural reinterpretation of his work within the framework of $Q$-cohomology. Finally, we explain how to use this method to prove dimensional reduction for more generic actions.

Let us start by considering the following deformed partition function with sources \cite{Cardy:1983aa},
\begin{equation}
   Z_{\lambda}[J]
   = \int \mathcal{D} \Phi e^{{- S_{\lambda}}-\int d^{d|2}y\,\delta_\perp J\Phi} \, ,
\end{equation}
where the source term  $\int d^{d|2}y\,\delta_\perp J\Phi $ is integrated in superspace with a delta function $\delta_\perp\equiv 2\pi \delta(\theta)\delta(\thetab)\delta(\x)\delta(\y)$ which projects the integrand to $\mathbb{R}^{d-2}$, and  $S_\lambda$ is the following deformed  action
\begin{equation}
    S_\lambda = \lambda S_{\text{SUSY}}+(1-\lambda)(\Sred+S_\perp)\, .
\end{equation}
The orthogonal action $S_\perp$ here is defined by splitting the superfield Lagrangian as $\mathcal{L}_{\text{SUSY}}=\mathcal{L}_{\|}+\mathcal{L}_{\perp}$, where
\begin{eqnarray}
    \label{eq:SPSPart}
\!\!\!\!\!\!\!\!\!\!\!\!   \mathcal{L}_{\|} = -\frac{1}{2} \Phi \partial^2_{\|} \Phi+V(\Phi) \, ,  \quad
   & &  S_{\|}=
    \int d^{d}x\,\bigl( \partial^i \omega \partial_i  \varphi+\omega V^{\prime}(\varphi)+\partial^i \psi \partial_i \bar{\psi}+\psi \bar{\psi} V^{\prime \prime}(\varphi) \bigr) \, ,
    \\
    \label{eq:SorthoPart}
 \!\!\!\!\!\! \!\!\!\!\!\! \mathcal{L}_{\perp} = -\frac{1}{2} \Phi \partial^2_{\perp} \Phi \, , \ \ \ \ \ \ \ \ \  \quad
  &&    S_\perp=\int d^{d}x\,\bigl( -\omega \partial_\x^2 \varphi-  \omega\partial_\y^2\varphi-\omega^2-\psi \partial_\x^2 \bar{\psi}- \psi \partial_\y^2 \bar{\psi} \bigr) \, .
\end{eqnarray}
The deformed action $S_\lambda$ defines a continuous interpolation between $S_{\text{SUSY}}$ (at $\lambda = 1$) and $\Sred+S_\perp$ (at $\lambda = 0$). 
Moreover, at $\lambda=0$, the action $S_\perp$ decouples from the rest of the terms.
Specifically, Cardy argues that the fields $\partial_\x\varphi,\partial_\y\varphi$ and $\omega$ inside $S_{\perp}$ decouple when computing correlators of $\varphi(x_\parallel)$ with the reduced action.
Thus, $Z_{\lambda}[J]$ defines an interpolation between the SUSY theory and the reduced one.
The idea is to prove that $Z_{\lambda}[J]$ is independent of $\lambda$, which thus implies dimensional reduction.

To prove this statement, it is convenient to rewrite $\Sred$ in terms of $\mathcal{L}_{\|}$:
\begin{equation}
    \qquad \Sred = 2\pi\int d^{d|2}y \, \delta_\perp \mathcal{L}_{\|}  \, .
\end{equation}
Then $\partial_\lambda \ln      Z_{\lambda}[J]$ can be expressed in terms of the following compact expression: 
\begin{equation}
    \partial_\lambda \ln      Z_{\lambda}[J]=- 2\pi\int d^{d|2} y\,(1-\delta_{\perp})\left\langle\mathcal{L}_{\|}(y)\right\rangle_{\lambda, J} \, ,
    \label{eq:dlnZ}
\end{equation}
which is computed with the deformed action $S_\lambda$.
Since the action $S_\lambda$, including the source term, is manifestly invariant under $OSp(2|2)$ super-rotations, it follows that $\mathcal{L}_{\|}$ is also invariant. Consequently, the one-point function above is constrained to take the following form
\begin{equation}
    2\pi \left\langle\mathcal{L}_{\|}(y)\right\rangle_{\lambda, J}=F_{\lambda, J}\left(x_\parallel, y_{\perp}^2\right) \, ,
    \label{eq:ConstrainedL}
\end{equation}
for some function $F_{\lambda, J}$. For brevity, we henceforth suppress the $\lambda, J$ subscripts and the dependence on $x_{\|}$. Substituting \eqref{eq:ConstrainedL} into \eqref{eq:dlnZ}, we get
\begin{equation}
   \partial_\lambda \ln      Z_{\lambda}[J]=-\int d^{d-2}x_{\|}d^{2|2}y_{\perp}\,F\left(y_{\perp}^{2}\right)+\int d^{d-2}x_{\|}\,F\left(0\right).
    \label{eq:dlnZ(F)}
\end{equation}
To compute the first integral, we expand $F_{\lambda, J}$ in the Grassmann variables $\theta\bar{\theta}$:
\begin{equation}
    F(r^2 - 2\theta\bar{\theta}) = F( r^2)-2 \theta \bar{\theta} \frac{\partial F(r^2)}{\partial r^2} \, ,
\end{equation}
with the usual notation $r^2=\x^2+\y^2$. Then the perpendicular part of the integral of $F(y_\perp^2)$ evaluates  to
\begin{equation}
\begin{aligned}
\label{F0-Finf}
    \int d^{2|2} y_\perp \, F(y_{\perp}^{2})
    & =-\frac{1}{\pi} \int r d r d \eta \frac{dF(r^2)}{d r^2}  =F(0)-F(\infty)\,.
\end{aligned}
\end{equation}
where $\eta$ is a polar angle in the $\x$-$\y$ plane. Moreover, $F(\infty)$ vanishes because of the standard boundary conditions of fields at infinity. Therefore plugging \eqref{F0-Finf} in \eqref{eq:dlnZ(F)}  we finally obtain the desired relation
\begin{equation}
 \partial_\lambda \ln      Z_{\lambda}[J]=  0\,.
\end{equation}
This shows that the generating functional $Z_\lambda$ is independent of $\lambda$. Thus, it does not matter whether we take $\lambda = 1$ or $\lambda = 0$. Therefore, the theory $S_{\text{SUSY}}$ (for $\lambda = 1$) dimensionally reduces to $\Sred$ (for $\lambda = 0$).

\subsection{Interpolation method from \(Q\)-cohomology} 

In this section, we reformulate Cardy’s argument using the cohomology of $Q$.

To make the argument nicer, we slightly change the definition of the deformed partition function as follows
\begin{equation}
   Z_\lambda
   = \int \mathcal{D} \Phi e^{{- S_{\text{SUSY}}+\lambda( S_{\text{SUSY}}-\Sred)}}\,.
\end{equation}
In this notation, the interpolation is even cleaner and gives on the nose $S_{\text{SUSY}}$ at $\lambda=0$ and $\Sred$ at $\lambda=1$.
Again, we want to prove that the deformation does not depend on $\lambda$, but now the strategy is to simply prove that the term $S_{\text{SUSY}}-\Sred$ is $Q$-exact (it is explicitly $Q$-closed since these terms are invariant under $OSp(2|2)$). 

We indeed find that the deformation is $Q$-exact:
\begin{equation}
\boxed{
    S_{\text{SUSY}}-\Sred = Q \int d^{d}x\,\bigl( \partial^{i}\psi_{\|}\partial_{i}\varphi+\psi_{\|}  V'(\varphi)+\omega \psi_\perp \bigr)  }\, ,
     \label{S-Sred=Qpar}
\end{equation}
where we defined two Grassmann fields $ \psi_{\|}$ and $ \psi_\perp$ as follows
\begin{equation}\label{psi-par-perp}
    \psi_{\|} \equiv \frac{1}{r^2}(\x\bar\psi+\y\psi) \, , 
    \qquad
    \psi_\perp \equiv -\partial_\y \psi - \partial_\x \bar \psi \, .
\end{equation}
In order to prove \eqref{S-Sred=Qpar}  we use the following transformations of the newly introduced Grassmann fields
\begin{equation}\label{Qpsi-par-perp}
Q\psi_{\|} = \omega + \frac{1}{r} \partial_r \varphi \,, \qquad Q \psi_\perp = -(\partial_\x^2+\partial_\y^2) \varphi - (r \partial_r +2)\omega \, ,
\end{equation}
which further give
\begin{align}
    \label{eq:S_parTerms}
Q  ( \partial^{i}\psi_{\|}\partial_{i}\varphi+\psi_{\|}  V'(\varphi)) &=\partial^{i}\omega\partial_i \varphi +\partial^{i}\psi \partial_{i}\bar{\psi}+ \omega V^{\prime}(\varphi)+\psi \bar{\psi} V^{\prime \prime}(\varphi)   \\
    \label{eq:red_parTerms}
&\quad +\frac{1}{2r}\partial_{r}(\partial^{i}\varphi\partial_{i}\varphi)
 + \frac{1}{r} \partial_r V(\varphi)
 \, .
\\
    \label{eq:bosoSector}
Q(\omega \psi_\perp) &= -\omega (\partial_\x^2+\partial_\y^2) \varphi- 2 \omega^2 - \omega \x \partial_\x\omega- \omega \y \partial_\y\omega    \\
    \label{eq:fermSector}
    &\quad+ \partial_{\x}\psi\partial_{\x}\bar{\psi}+\partial_{\y}\psi\partial_{\y}\bar{\psi}+\partial_{\x}\bar{\psi}\partial_{\y}\bar{\psi}+\partial_{\x}\psi\partial_{\y}\psi\,.
\end{align}

Notice that the terms in \eqref{eq:S_parTerms} integrate to $S_{\parallel}$ while the ones 
 in \eqref{eq:red_parTerms} integrate to minus the reduced action, namely $ \int d^{d}x\,\bigl(\frac{1}{2r}\partial_{r}(\partial^{i}\varphi\partial_{i}\varphi)
 + \frac{1}{r} \partial_r V(\varphi)\bigr)=-S_{d-2}$, which can be easily seen by following the same logic as \eqref{eq:SAlmostRedu}: one writes the measure $d\x d\y=rdrd\eta$ and integrates radially to get a (vanishing) boundary term at $r=+\infty$ and an $r=0$ contribution which has no $\eta$ dependence.
We can also easily show that the integral of $Q(\omega \psi_\perp) $ is equal to the orthogonal action $\eqref{eq:SorthoPart}$. To see this we first notice that $-\omega \x \partial_\x\omega$ has the same integral as $\omega^2/2$ by integration by parts.
Moreover we notice that $\psi$ and $\partial_\x \partial_\y \psi$ anticommute, which implies the identity $2\,\partial_\x \psi\, \partial_\y \psi = \partial_\x(\psi\,\partial_\y \psi) + \partial_\y(\partial_\x \psi\, \psi)$, showing that the last two terms in \eqref{eq:fermSector} are total derivatives. 

Actually, the argument above shows a bit more than \eqref{S-Sred=Qpar}. Specifically we found that the orthogonal action $S_{\perp}$ and the difference $(S_{\parallel}-S_{d-2})$ are separately $Q$-exact. 
This can also be used to rephrase the original argument of Cardy. There one needed to argue both that $S_{\perp}$ decoupled and that $(S_{\parallel}-\Sred)$ did not modify the action. 
Now we can explain that both terms are in fact $Q$-exact. 

Finally, let us stress that we did not write the source terms for brevity.
However, as explained in the review \autoref{sec:loc_review}, one can always add $Q$-invariant sources without changing the argument. 
This means that we can localize all correlation functions of operators in the cohomology of $Q$. These were described in detail in \autoref{sec:cohomology} and were shown to match the set of operators of the reduced theory. 
We therefore conclude that the interpolation method provides a very simple and clean argument for dimensional reduction.

\subsection{Generalizing to other theories} 
\label{sec:gen_Cardy}
In this section, we argue that Cardy's argument is, in fact, very generic and can be applied to a large class of different theories. 
Given a generic action $S$ invariant under PS SUSY, we can prove its dimensional reduction by pursuing the following strategy:
\begin{itemize}
    \item We define $S=S_\parallel+S_\perp$, where 
    $S_\parallel \equiv \int d^{d|2}y \, \mathcal{L}_{\parallel} $ 
    is defined by  setting the orthogonal part of the metric to zero inside $S$.
    \item By definition $S_\perp$ is an $OSp(2|2)$-singlet built out of the metric $g_{2|2}$. 
    By the decoupling argument of \eqref{def:decoupling_KRT}, which is valid for any model, we thus conclude that $S_\perp$ decouples. For scalar actions we can further show that this term is indeed $Q$-exact, as explained below.
    \item  Any deformation of the action by $S_\parallel-S_{d-2}$ does not change the path integral, where we define $\Sred \equiv \int d^{d-2} x_{\|} \, \mathcal{L}_{\parallel}|_{d-2}$.
    This is automatically proven by the original argument by Cardy \eqref{eq:dlnZ(F)}, which was agnostic on the form of the actions. Alternatively, below we will also show that any action $S_\parallel$  is indeed equal to $S_{d-2}$ in cohomology.  
    \item Following Cardy's argument, the deformed action $ S - \lambda (S-\Sred) $ does not change the partition function and interpolates between $S$ and $\Sred$ by tuning $\lambda$ from $0$ to $1$.
\end{itemize}
We thus obtained that any PS supersymmetric action $S$ dimensionally reduces to $\Sred$. 
In this paper, we focus on scalar actions built out of the scalar superfield $\Phi$, e.g., in appendix~\ref{App:proof_Q_exact} we detail how to use this algorithm in the case of a higher-derivative term $\Phi (\partial^2)^n \Phi$.
It is, however, important to stress that the same strategy can also be extended to more general models, which might also contain gauge fields and fermions. It would be interesting to spell out the details of such cases, but we leave it for future work.

In the following, we want to show that, for any $S$, both $S_\perp$ and  $S_\parallel-S_{d-2}$ are $Q$-exact. To do so we first prove that for any scalar function $F$ of the superfields, the following equation holds
\begin{equation}
\label{Fh-Fl=Q}
\boxed{
 \int  d^{2|2}y_{\perp} \, F(y)  =  F(x_\parallel,y_{\perp}=0)+ Q\int d^{2|2}y_{\perp} \ \chi \,  F(y) } \, ,
\end{equation}
 where we recall that $d^{2|2}y_{\perp}=d\x \, d\y \,  [d\bar{\theta}][d\theta]$ and we introduced the Grassmann variable $\chi$ defined as
\begin{equation}
\label{def:csi}
 \chi \equiv \frac{ 
 (\y \theta - \x \thetab)}{r^2} \, .
\end{equation}
Equation \eqref{Fh-Fl=Q} can be equivalently rewritten as
\begin{equation}
\label{Fh-Fl=Q1}
 \frac{1}{2\pi} \int  d\x \, d\y \, F_{\text{highest}}(x)  =  F_{\text{lowest}}(x_\parallel,\x=0,\y=0)+ Q\int d^{2|2}y_{\perp} \ \chi \,  F \, , 
\end{equation}
where $F_{\text{highest}}$ ($F_{\text{lowest}}$)  is the highest (lowest) component of $F$ when expanded in $\theta$, $\thetab$. This form of the equation makes it manifest that the lowest and highest components of superfields are related in cohomology.   Very similar identifications of lowest components to integrated highest components of chiral multiplets (in $Q$-cohomology) were observed already for 2d ${\cal N}=(2,2)$ and 4d ${\cal N}=2$ theories on spheres, and were used to relate K\"ahler potentials on superconformal manifolds to sphere partition functions~\cite{Gerchkovitz:2014gta,Gomis:2014woa} (see also \cite{Gerchkovitz:2016gxx,Ishtiaque:2017trm}).

First, let us show how to prove \eqref{Fh-Fl=Q}. We start by performing the following computation 
\begin{align}
\label{Qpar_rel}
    Q\int [d\bar{\theta}][d\theta] \, \chi F
    = - \int [d\bar{\theta}][d\theta] \, \chi q F
    = \int [d\bar{\theta}][d\theta]\left(\frac{\x^{2}\bar{\theta}\partial_{\bar{\theta}}+\y^{2}\theta\partial_{\theta}}{r^{2}}+\theta\bar{\theta}\frac{1}{r}\partial_{r}\right) F \, ,
\end{align}
where in the first equality we use that  $F$ is a scalar, so the action of $Q$ is replaced by the Killing vector $q$, while the second equality is obtained by replacing the form of the Killing vector $q$ and simplifying the resulting differential operator.

Next, we simplify the right-hand side of \eqref{Qpar_rel}. Integrating by
parts  $\theta\,\partial_{\theta}$ and
$\bar{\theta}\,\partial_{\bar{\theta}}$, the first term in parentheses reduces to one. 
We then perform the Grassmann integration of the term
proportional to $\theta\bar{\theta}$, which selects the lowest component
of $F$. Altogether, we find
\begin{equation}
\label{Q=Fhigh-Fred}
    Q\int [d\bar{\theta}][d\theta] \, \chi F=\int [d\bar{\theta}][d\theta] F +\frac{1}{2\pi} \frac{1}{r}\partial_{r}F_{\text{lowest}} \,. 
\end{equation}
Finally, \eqref{Fh-Fl=Q} is obtained by integrating this expression using polar coordinates in $\mathbb{R}^2$ and noticing that $\frac{1}{r}\partial_{r}F_{\text{lowest}}$ becomes a total derivative that picks the boundary term at $r=0$.

Equation \eqref{Fh-Fl=Q} is very powerful since it holds for any scalar superfield~$F$.
First we can use \eqref{Fh-Fl=Q} to prove that any $S_\perp[\Phi]$ is $Q$-exact by itself. Indeed, by definition $S_\perp$ must be defined as 
\begin{equation}
\label{S_perp_gen}
    S_\perp= \int d^{d|2}y \, \Ocal(y) = \int d^{d-2}x_\parallel \int d^{2|2}y_{\perp} \; \mathcal{O}(y) \, ,
\end{equation}
for some scalar superfield operator $\Ocal$ built out of the orthogonal metric $g_{2|2}$. 

Moreover, following the logic of the equation \eqref{def:singlets}, the most generic form of these superfield operators can be written as 
\begin{equation}
    \Ocal(y) \equiv  \lim_{y_i \to y} P_\parallel P_\perp \Phi(y_1) \cdots \Phi(y_n)  \, ,
\end{equation}
where $P_\perp $ must be non-trivial and $P_\parallel$ here is defined by taking $P_\parallel^{\{\mu\}}$ in \eqref{def:singlets} and contracting all the $\mu$ indices to generate a scalar operator. Using \eqref{Fh-Fl=Q} we can thus argue that
\eqref{S_perp_gen} is equal in cohomology to $\int d^{d-2} x_{\|} \, \Ocal(y)|_{d-2}$, which is exactly the integral of one of the operators in \eqref{def:singlets}, which was proven to be $Q$-exact. This thus shows that $S_\perp$ must also be  $Q$-exact and thus that it decouples.

We can also use  \eqref{Fh-Fl=Q} to show that 
any parallel superfield action $S_\parallel=\int d^{d|2}y\, \mathcal{L}_{\parallel} $ is equal in cohomology to  a reduced action which can be simply defined as $\Sred=\int d^{d-2} x_{\|}\, \mathcal{L}_{\parallel}|_{d-2}$. 
This trivially arises by using \eqref{Fh-Fl=Q} with $F=\int d^{d-2} x_{\|}\, \mathcal{L}_{\parallel}$ for any $\mathcal{L}_{\parallel}$,
namely
\begin{equation}
    S_{\|}-\Sred= Q \int d^{d|2}y \, \chi \,  \mathcal{L}_{\|} \, .
    \label{eq:GenericEq}
\end{equation}

Interestingly, we can also apply \eqref{Fh-Fl=Q} to probe the dimensional reduction of extended defect operators. To this end, we define $F=\int d^{p-2} x_{\|} \, \mathcal{O}(y)$, where a superfield operator $\mathcal{O}$ is integrated
over a bosonic submanifold  $\mathbb{R}^{p-2} \subset \mathbb{R}^{d-2} $.  Then~\eqref{Fh-Fl=Q} implies that
\begin{equation}
    \int d^{p|2} y \, \mathcal{O}(y) =  \int d^{p-2} x_\parallel \, \mathcal{O}(x_\parallel, y_\perp\!=0) + Q[{\dots}] \, ,
\end{equation}
which means that there are two defects which are equal up to $Q$-exact terms. 
This setup was recently studied in  \cite{Ghosh:2025lzb}, where indeed it was found that these two defects have the same dimensional reduction. 
Finally, let us mention that to derive \eqref{Fh-Fl=Q} we did not have to specify that $F$ was a function of scalar superfields $\Phi$, so it should also work for gauge superfields and fermions. It would be interesting to consider these setups. 

Another interesting application concerns the reduction of conserved charges.
As reviewed in appendix~\ref{sec: App A}, the topological charge
associated with a supercurrent $\mathcal{J}^{a}$ can be written as an integral
over a codimension-$1$ submanifold $\Sigma$ of superspace,
\begin{equation}
\int_{\Sigma} dS^{a} \, \mathcal{J}_{a}(y) \, ,
\end{equation}
where $dS^{a}$ is the integration measure induced from $d^{d|2}y$.
When $\Sigma=\Sigma_{d-3}\times\mathbb{R}^{2|2}$ is a product, in which $\Sigma_{d-3}$ is a hypersurface in $\mathbb{R}^{d-2}$ (such as a time slice in any choice of foliation), the relevant components of $\mathcal{J}_{a}$ are $\mathcal{J}_i$, $i=1,\dots,d-2$.
Applying \eqref{Fh-Fl=Q} to these superfields $\mathcal{J}_i(y)$, we find that the charge above reduces to
\begin{equation}
    \int_{\Sigma} dS^{a} \, \mathcal{J}_{a}(y)
    = \int_{\Sigma_{d-3}} dS_{d-3}^i \int d^{2|2}y_{\perp} \, \mathcal{J}_{i}(y)
    = \int_{\Sigma_{d-3}} dS_{d-3}^i \, \mathcal{J}_{i}(x_{\|}, y_{\perp}=0) + Q[{\cdots}] \, .
\end{equation}
Passing to $Q$-cohomology, this is the conserved charge of the dimensionally reduced theory.
It is worth noting that this reduction does not apply to all conserved
charges. For example, for supertranslations $P^{a}$,   it only applies to the components  $P^{i}$ for $i=1,\dots,d-2$ because
\eqref{Fh-Fl=Q} relies on the $OSp(2|2)$ invariance of the integrand.
The reduction charges should be understood as an operator equation valid inside correlation functions of $Q$-closed operators, or as equations in the dimensionally reduced theory.

\section{Discussion} \label{sec:conclusions}
In this paper, we revisited dimensional reduction in Parisi--Sourlas (PS)
supersymmetric theories. We provided three arguments based on the
cohomology of the supercharge $Q$ defined in \eqref{def:Q}.

First, we reformulated the decoupling argument of \cite{paperI} in
cohomological terms. In particular, in \cite{paperI} it was shown that
the operators in the reduced theory must arise from $OSp(2|2)$-singlets,
and that all singlets built out of the orthogonal metric $g_{2|2}$ must
decouple. The $OSp(2|2)$-singlets are, in particular, $Q$-closed, and in
section~\ref{sec:cohomology} we proved that operators built out of
$g_{2|2}$ are $Q$-exact. Therefore, we showed that the set of operators
of the dimensionally reduced theory is defined by $OSp(2|2)$-singlets
in the cohomology of $Q$. This is not only aesthetically pleasing, but
also parallels familiar cohomological constructions in other
supersymmetric contexts, see e.g.~\cite{Beem:2013sza}.

Second, we presented a localization argument based on a canonical
$Q$-exact deformation, phrased in a way that aligns with the standard
localization logic in high-energy physics, see e.g.~\cite{Pestun:2016zxk}.
We also explained how this setup differs from the earlier approach of
Zaboronsky \cite{Zaboronsky:1996qn}, where a non-standard deformation
leads to additional subtleties.

Third, we revisited Cardy's interpolation argument, where one considers
a generating functional for the reduced correlators written in terms of
a deformed action which interpolates between the supersymmetric and the
reduced theories as $S_{\text{SUSY}} - \lambda \left( S_{\text{SUSY}} - S_{d-2} \right)$.
Dimensional reduction is then established by showing that the
generating functional is independent of $\lambda$. Part of Cardy's
original argument was completely generic and relied only on PS
supersymmetry, while another part depended on the explicit form of
$S_{\text{SUSY}}$, which was taken to be a supersymmetric scalar action
with a generic potential. In section~\ref{sec:Cardy} we showed that
$\left( S_{\text{SUSY}} - S_{d-2} \right)$ is in fact a $Q$-exact
deformation, which explains why it can be added with an arbitrary
coefficient without affecting the physics. Moreover, we demonstrated
that the interpolation argument can be generalized to any action with
PS supersymmetry, including setups involving extended operators.

Altogether, these three arguments fill an important gap in the
literature by clarifying how various proofs of dimensional reduction
are not isolated results, but can instead be organized within a
unified $Q$-cohomological framework. This perspective also provides a
natural bridge to the language commonly used in the study of standard
supersymmetry in high-energy physics.

Some of the techniques employed in the study of PS SUSY—such as the decoupling argument of \cite{paperI}, localization à la Zaboronsky \cite{Zaboronsky:1996qn}, and the interpolation method introduced by Cardy \cite{Cardy:1983aa}—are largely unfamiliar to high-energy physicists working on conventional supersymmetry. We hope that our reformulation in terms of the cohomology of a supercharge may help uncover new applications of these methods also in that broader context.
It is worth emphasizing that some of these techniques are remarkably powerful and general.
For instance, the argument of \cite{paperI} appears to apply to any theory that localizes from a superspace onto a submanifold, provided that the component of the superspace metric orthogonal to the submanifold is supertraceless. 
It would be interesting to explore whether this type of reasoning could be extended to other models, such as \cite{Beem:2013sza}, providing an alternative perspective on why certain sectors of operators effectively localize to lower dimensions. Similarly, it would be very interesting to investigate whether Cardy’s interpolation method can be applied in more standard supersymmetric settings. This approach is considerably simpler than conventional localization, as it does not require the computation of one-loop determinants, and is therefore particularly well suited for non-compact situations.

Beyond applications to models with conventional supersymmetry, it would also be interesting to further investigate systems with PS SUSY\@. Most existing studies have focused on scalar theories; however, PS-supersymmetric QFTs can also be constructed with gauge fields and fermions, see e.g. \cite{GOZZI1989356, Kondo:1998sr, Kalkkinen:1996fg, Magpantay:1999xi, PhysRevD.110.034505, MCCLAIN1983430}. As shown in section~\ref{sec:Cardy}, dimensional reduction holds for all PS SUSY QFTs. Although the intrinsic non-unitarity of these models prevents their use as fundamental descriptions of particle interactions, dimensional reduction makes them valuable as theoretical tools or toy models: it provides a controlled setting in which higher-dimensional observables can be accessed through the simpler kinematics of a lower-dimensional effective description.

The generality of dimensional reduction also has important implications for dimensional uplift. Generic PS SUSY actions reduce to fairly generic lower-dimensional theories, which in turn implies that a large class of lower-dimensional models admit an uplift. The abundance of such uplifts provides evidence that dimensional uplift may hold generically, at least for Lagrangian models. It would be interesting to determine whether counterexamples exist within this class. We leave this question to future work.

It is also worth noting that a version of PS dimensional reduction has
been observed in holographic correlators for various SCFTs
\cite{Behan:2021pzk, Alday:2021odx}. In that context, however, no clear
explanation of this behavior was provided. Studying PS models with more
general field content might therefore help clarify whether certain
sectors of these SCFTs admit an effective description in terms of PS
supersymmetry.

A natural further direction is to extend these ideas beyond flat space.
In particular, it would be interesting to investigate whether
dimensional reduction persists for PS-supersymmetric theories defined
on curved backgrounds, such as AdS or dS.

It is also tempting to explore whether similar constructions can be
applied to gravitational models. In this context, one could study
gravitational actions invariant under PS supersymmetry (see
\cite{Kellett:2020rjw} for a construction in this direction) and
understand under which conditions such theories undergo dimensional
reduction. More generally, this framework could provide a way to
engineer or isolate sectors of gravitational theories that effectively
behave as lower-dimensional systems.

By framing dimensional reduction in terms of $Q$-cohomology and
extending it to general PS-supersymmetric theories, we hope to provide a
unifying viewpoint that will both clarify the structure of PS models and
inform applications in high-energy physics.

\acknowledgments
BLF thanks Masahito Yamazaki for useful discussions on one-loop determinants.
GP thanks Weam Abou Hamdan, Soumangsu Chakraborty, Guido Festuccia, Pierre Heidmann, Kakha Shamanauri, Suthanth Srinivasan, and Maxim Zabzine for useful discussions. He is also grateful to the organizers and participants of the 22nd Simons Physics Summer Workshop and the 50 Years of the Black Hole Information Paradox program at the Simons Center for Geometry and Physics, where part of this work was done.
ET thanks
Slava Rychkov,  Apratim Kaviraj, Balt  van Rees for useful discussions. 
He also thanks the Yukawa Institute for Theoretical Physics at Kyoto University for the kind hospitality during the workshop “Progress of Theoretical Bootstrap”.
ET is partially supported by HORIZON-MSCA-2023-SE-01-101182937-HeI\@.
GP is supported by the Department of Physics at The Ohio State University.
BLF is partly supported by the Projet-ANR-23-CE40-0010 and HORIZON-MSCA-2022-SE/101131233 grants.

\appendix
\section{Conventions and useful identities}
\label{sec: App A}
In this appendix, we set the conventions about charges, generators of transformations, and algebra, and derive useful identities. 


Supertranslation and superrotation charges are obtained by appropriately integrating the superstress tensor $\mathcal{T}_{ab}$ with a 
a vector field $\epsilon^{a}(y)$,
\begin{equation}
  Q_{\epsilon}(\Sigma)
  = - \int_{\Sigma} dS^{a}\, \mathcal{J}^{(\epsilon)}_{a}(y)
  \equiv - \int_{\Sigma} dS^{a}\, \mathcal{T}_{ab}(y)\,  \epsilon^{b}(y)\,,
  \label{def:TopCharge}
\end{equation}
where $\Sigma$ is a codimension-one hypersurface in superspace. The operator $Q_{\epsilon}(\Sigma)$ is topological as long as the supercurrent $\mathcal{J}^{(\epsilon)}_{a}$ is conserved.
The conservation of  $\mathcal{J}^{(\epsilon)}_{a}$ leads to the Killing equation 
$\partial^a \epsilon^{b}+\partial^b \epsilon^{a}=0$,
which admits the solution
\begin{equation}
  \epsilon^{a} = \omega^{a}+\omega^{ab}\,  y_{b}\,,
\end{equation}
where  $\omega^{a}$ is the vector of parameters associated to supertranslations and $\omega^{ab}$ are the 
 transformation parameters for superrotations, which  transform in the graded antisymmetric representation,
$\omega^{ab} = -(-1)^{[a][b]}\, \omega^{ba}$.
The Killing vectors are then defined as $\epsilon=\epsilon^a\partial_a$ and can be expanded in the basis of parameters as follows,
\begin{equation}
\label{eq:2KillingSols}
  p^a = \partial^a, 
  \qquad
  m^{ab} = y^{a}\partial^{b} - (-1)^{[a][b]} y^{b}\partial^{a}\,.
\end{equation}

Let us now review how the action of the charge $Q_{\epsilon}$ on a scalar superfield can be computed using the Killing vectors.
We make use of the Ward identity for the current $\mathcal{J}^{(\epsilon)}_{a}$:
\begin{equation}
\partial_y^{a}
  \left\langle
   \mathcal{J}^{(\epsilon)}_{a}(y)\,\Phi(y_{1})\cdots\Phi(y_{n})
  \right\rangle
  =
  - \sum_{i=1}^{n} \delta \big(y - y_{i}\big)\,
  \epsilon(y_i)\left\langle
    \Phi(y_{1})\cdots\Phi(y_{n})
  \right\rangle .
\end{equation}
Let us surround a given insertion $\Phi(y_{i})$ by a small ball $B_{i}$ whose
boundary $\Sigma_{i}$ does not enclose any other operator insertions.
Integrating the Ward identity over $B_{i}$ and using the divergence theorem,
we obtain
\begin{align}
\label{action_of_charges}
  \int_{B_{i}} d^{d|2}y\,
  \partial_y^{a}
  \bigl\langle
   \mathcal{J}^{(\epsilon)}_{a}(y)\,\Phi(y_{1})\cdots\Phi(y_{n})
  \bigr\rangle
  &=
  \int_{\Sigma_{i}} dS_{a}\,
  \bigl\langle
    \mathcal{J}^{a}(y)\,\Phi(y_{1})\cdots\Phi(y_{n})
  \bigr\rangle \nonumber \\
  &=
  - \epsilon(y_i)\langle
      \Phi(y_{1})\cdots \Phi(y_{n})
    \rangle \, .
\end{align} 
Using \eqref{def:TopCharge}, this implies
\begin{equation}
  \left\langle
    Q_{\epsilon}(\Sigma_{i})\,\Phi(y_i)\cdots
  \right\rangle
  =
  \left\langle \epsilon(y_i)\Phi(y_{i})\cdots
  \right\rangle .
\end{equation}
Or, as an operator equation,
\begin{eqnarray}
    Q_{\epsilon}\, \Phi =\epsilon \, \Phi
\, .
\end{eqnarray}
When quantizing the theory, this relation is promoted to a
commutator, $[Q_\epsilon, \Phi\} = \epsilon\,\Phi\,$.
The charge can also be expanded on the basis of parameters  $Q_\epsilon=\omega^a P_a+\omega^{ab}M_{ab}$, where $P_a$ and $M_{ab}$ respectively define the generators of supertranslations and superrotations in accordance with \eqref{eq:2KillingSols}.

We will often use the action of a charge on the components of a superfield.
Let us exemplify, for the charge $P_\theta$, how to compute it,
\begin{align}
\label{eq:pPhi}
    P_\theta \Phi(y) &=p_\theta \Phi(y) = \bar{\psi}(x) + \bar{\theta}\omega(x) \, ,
    \\
    &= P_\theta\varphi(x) -\theta P_\theta \bar\psi(x)-\thetab P_\theta \psi(x) + \theta\thetab P_\theta \omega(x) \, ,
    \label{Ptheta_line2}
\end{align}
where the minus signs in \eqref{Ptheta_line2} appear because $P_\theta$ anticommutes with Grassmann numbers. From here, we deduce 
\begin{equation}
\begin{gathered}
    P_\theta\varphi = \bar{\psi},\quad P_\theta \bar\psi = 0, \quad
   P_\theta \psi = -\omega,\quad P_\theta \omega= 0 \, . 
\end{gathered}
\end{equation}
One can easily check that this variation, and its analogues for other supertranslations and superrotations, preserve the action~\eqref{eq:PS_action}.

Let us now consider the repeated action of $Q_\epsilon$ on a scalar superfield.
We find that the repeated action reverses the order of the differential operators,
\begin{equation}
  Q_{\epsilon_1}Q_{\epsilon_2}\Phi
  = (-1)^{[\epsilon_1][\epsilon_2]}\, \epsilon_2 \, \epsilon_1 \, \Phi 
  \, .
  \label{eq:Q1Q2}
\end{equation}
To see this, we first write $Q_{\epsilon_2}\Phi=\epsilon_2\Phi$, then use that the action of the charge $Q_{\epsilon_1}$ 
graded commutes with the differential operator $\epsilon_2$ since it only acts on
$\Phi$, in accordance with  \eqref{action_of_charges}. 
As a simple example, consider the translation generator $P_\theta$ and the
rotation generator $M^{\x\theta}$, with differential representations
$p_\theta$ and $m^{\x\theta}$, respectively. Then
\begin{equation}
  P_\theta M^{\x\theta}\Phi
  = -m^{\x\theta} p_\theta \Phi
  = -\x\,\omega
    + \theta\,\partial_{\x}\bar{\psi}
    + \theta\bar{\theta}\,\partial_{\x}\omega \, .
\end{equation}
To see how the algebra of charges is related to that of the Killing vectors, we use \eqref{eq:Q1Q2} to write 
\begin{align}
    [Q_{\epsilon_{1}},Q_{\epsilon_{2}}\}\Phi&=Q_{\epsilon_{1}}Q_{\epsilon_{2}}\Phi-(-1)^{[\epsilon_{1}][\epsilon_{2}]}Q_{\epsilon_{2}}Q_{\epsilon_{1}}\Phi\\
    &=(-1)^{[\epsilon_{1}][\epsilon_{2}]}\epsilon_{2}\epsilon_{1}\Phi-\epsilon_{1}\epsilon_{2}\Phi \, ,
\end{align}
from which we find
\begin{equation}
    \left[Q_{\epsilon_1}, Q_{\epsilon_2}\right\}=Q_{-\left[\epsilon_1, \epsilon_2\right\}}
    \, .
\end{equation}

In the paper, we focus on a specific fermionic charge denoted by $Q$ given by the following combination of superrotations:
\begin{equation}
\label{def:Qapp}
    Q \equiv M^{\x\theta} + M^{\y\bar{\theta}} ,
\end{equation}
and we denote the corresponding differential operator as $q$, explicitly,
\begin{equation}
    q= m^{\x\theta}+m^{\y\thetab} = \x \partial_{\bar{\theta}}-\theta \partial_{\x}-\y \partial_\theta-\bar{\theta} \partial_{\y} \, .
\end{equation}
The superrotation generators $M ^{ab}$ satisfy the algebra \eqref{algebra_M_OSp}, but,
given the definition of $Q$, it is useful to spell out the anticommutators of the fermionic generators
\begin{equation}
    \{M^{\mu p}, M^{\nu q}\} = -g^{pq} M^{\mu \nu} + g^{\mu \nu} M^{pq} \, ,
\end{equation}
where $p, q = \theta , \bar{\theta}$. To be more explicit,  
\begin{equation}
\begin{aligned}
    &\{M^{\x\theta}, M^{\x\theta}\} = M^{\theta\theta}, \qquad 
    \{M^{\x\theta}, M^{\x\bar{\theta}}\} = M^{\theta\bar{\theta}}, \\
    &\{M^{\x\theta}, M^{\y\bar{\theta}}\} = M^{\x\y}, \qquad 
    \{M^{\x\theta}, M^{\y\theta}\} = 0 \, .
\end{aligned}   
    \label{eq:acommuRels}
\end{equation}
Using these relations, for the square of the charge, we obtain  
\begin{align}
    Q^2=\frac{1}{2}\{Q, Q\} & =\frac{1}{2}\{M^{\x \theta}, M^{\x \theta}\}+\frac{1}{2}\{M^{\x \theta}, M^{\y \bar{\theta}}\}+\frac{1}{2}\{M^{\y \bar{\theta}}, M^{\x \theta}\}+\frac{1}{2}\{M^{\y \bar{\theta}}, M^{\y \bar{\theta}}\} \nonumber\\
& = \frac{1}{2}(M^{\theta \theta}+M^{\bar{\theta} \bar{\theta}})+M^{\x\y} \, .
\end{align}

It is also convenient to compute the action of $Q$ on the components of $\Phi$, which can be obtained as follows,
\begin{align}
    Q \Phi & =(\x \partial_{\bar{\theta}}-\theta \partial_{\x}-\y \partial_\theta-\bar{\theta} \partial_{\y})(\varphi+\theta \bar{\psi}+\bar{\theta} \psi+\theta \bar{\theta} \omega) \nonumber \\
    & =\x \psi-\y \bar{\psi}+\theta\left(-\partial_{\x} \varphi-\x \omega\right)+\bar{\theta}\left(-\partial_{\y} \varphi-\y \omega\right)+\theta \bar{\theta}(\partial_{\y} \bar{\psi}-\partial_{\x} \psi)\\
    & = Q\varphi-\theta Q\bar{\psi}-\bar{\theta}Q \psi+\theta \bar{\theta} Q\omega \, .
\label{eq:QPhi}
\end{align}
where, again, the minus signs reflect the graded commutation of the fermionic charge with the
Grassmann coordinates. We can then recognize the variations of components that we wrote in the first line of \eqref{QPhi and Q^2Phi}.  Acting on these relations with $Q$ one more time, we can find the components of $Q^2\Phi$ reported in the second line of \eqref{QPhi and Q^2Phi}. 

The variation of components is especially useful when checking $Q$-invariance of terms expressed in components whose superfield form is unknown, such as the localization term \eqref{S_loc_explicit} in the action; see appendix \ref{sec: App B}.

\section{Comments on cohomology}
\label{App:proof_Q_exact}
\subsection{\(Q\)-exactness of operators built out of \texorpdfstring{$g_{2|2}$}{g(2|2)}}
\label{App:proof_Q_exact-subsec}
In this appendix, we provide a proof of equation~\eqref{eq:QexactBeast}, which can be written in the equivalent form
\begin{equation}
    (P_\perp F_\Phi)\bigr|_{d-2}
    = Q\Bigl[\bigl((\partial_{\theta_i}\partial_{\x_j}
        + \partial_{\thetab_j}\partial_{\y_i}) P'_\perp F_\Phi\bigr)\bigr|_{d-2} \Bigr] \,,
    \label{eq:CoolestEq}
\end{equation}
where \(P_\perp\) is defined in~\eqref{eq:Pperp} and is related to \(P'_\perp\) by
\(P_\perp = (\partial_{\perp i}\cdot \partial_{\perp j})\,P'_\perp\), and we have introduced the shorthand
\begin{equation}
    F_\Phi \equiv \Phi(y_1)\cdots\Phi(y_n)\,.
\end{equation}
We begin the proof by noting that, since the charge $Q$ does not act on the derivatives in front of $F_\Phi$ nor the restriction to $\mathbb{R}^{d-2}$, we may move it past them and let it act directly on $F_\Phi$:
\begin{align}
    Q\Bigl[\bigl((\partial_{\theta_i}\partial_{\x_j}
        + \partial_{\thetab_j}\partial_{\y_i}) P'_\perp F_\Phi\bigr)\bigr|_{d-2} \Bigr] &= - \bigl[(\partial_{\theta_{i}}\partial_{\x_{j}}+\partial_{\bar{\theta}_{i}}\partial_{\y_{j}})P_\perp 'QF_\Phi \bigr] \bigr|_{d-2} \nonumber
    \\& 
    = - \bigl[ P'_\perp(\partial_{\theta_{i}}\partial_{\x_{j}}+\partial_{\bar{\theta}_{i}}\partial_{\y_{j}})(q_{1}+\ldots+q_{n})F_\Phi \bigr]\bigr|_{d-2} \, ,
    \label{eq:MainEQ}
\end{align}
where $q_i=m^{\x_i\theta_i}+m^{\y_i\bar{\theta}_i}$, and we have moved $P'_\perp$ to the front (using that the derivatives $\partial_a$ graded-commute) for later convenience. 
At this point, it is useful to understand how the differential operator
\((\partial_{\theta_i}\partial_{\x_j} + \partial_{\bar{\theta}_i}\partial_{\y_j})\) acts on the combination
\(q\equiv q_1 + \cdots + q_n\). We have the anticommutators
\begin{equation*}
    \begin{aligned}
    \{\partial_{\theta_{i}}\partial_{\x_{j}}+\partial_{\bar{\theta}_{i}}\partial_{\y_{j}},q_{i}\} & = \{\partial_{\theta_{i}}\partial_{\x_{j}}+\partial_{\bar{\theta}_{i}}\partial_{\y_{j}}, \ \x_{i}\partial_{\bar{\theta}_{i}}-\theta_{i}\partial_{\x_{i}}-\y_{i}\partial_{\theta_{i}}-\bar{\theta}_{i}\partial_{\y_{i}}\}
    = -\partial_{\x_{j}}\partial_{\x_{i}}-\partial_{\y_{j}}\partial_{\y_{i}} ,
    \\
    \{\partial_{\theta_{i}}\partial_{\x_{j}}+\partial_{\bar{\theta}_{i}}\partial_{\y_{j}},q_{j}\} & = \{\partial_{\theta_{i}}\partial_{\x_{j}}+\partial_{\bar{\theta}_{i}}\partial_{\y_{j}}, \ \x_{j}\partial_{\bar{\theta}_{j}}-\theta_{j}\partial_{\x_{j}}-\y_{j}\partial_{\theta_{j}}-\bar{\theta}_{j}\partial_{\y_{j}}\}
    = \partial_{\theta_{i}} \partial_{\bar{\theta}_{j}} - \partial_{\bar{\theta}_{i}} \partial_{\theta_{j}} ,
    \end{aligned}
\end{equation*}
and the remaining $q_{k}$ anti-commute with $(\partial_{\theta_{i}}\partial_{\x_{j}}+\partial_{\bar{\theta}_{i}}\partial_{\y_{j}})$ trivially, so one gets
\begin{equation}
    (\partial_{\theta_{i}}\partial_{\x_{j}}+\partial_{\bar{\theta}_{i}}\partial_{\y_{j}})q=-\partial_{\perp i}\cdot \partial_{\perp j}-q(\partial_{\theta_{i}}\partial_{\x_{j}}+\partial_{\bar{\theta}_{i}}\partial_{\y_{j}}) \, .
    \label{eq:Obs1}
\end{equation}
Substituting the identity \eqref{eq:Obs1} into \eqref{eq:MainEQ}, we can rewrite the right-hand side as
\begin{equation}
\begin{aligned}
    \eqref{eq:MainEQ}
    &= (P_\perp F_\Phi)\bigr|_{d-2}
    + \bigl(P'_\perp q\,(\partial_{\theta_i}\partial_{\x_j}
      + \partial_{\bar{\theta}_i}\partial_{\y_j}) F_\Phi\bigr) \bigr|_{d-2}\, ,
      \label{eq:almostThere}
\end{aligned}
\end{equation}
where we observe that the first term coincides with the left-hand side of \eqref{eq:CoolestEq}. Therefore, to complete the proof, it suffices to prove that the second term in \eqref{eq:almostThere} vanishes. 
To show this, a key ingredient is the identity
\begin{equation}
    (\partial_{\perp i}\cdot \partial_{\perp j})(q_{i}+q_{j})=(q_{i}+q_{j})(\partial_{\perp i}\cdot \partial_{\perp j}) \, .
    \label{eq:Obs2}
\end{equation}
We can explicitly see that  this identity holds by using $
    \partial_{\perp i}\cdot \partial_{\perp j} = \partial_{\x_{i}}\partial_{\x_{j}}+\partial_{\y_{i}}\partial_{\y_{j}}+\partial_{\bar{\theta}_{i}}\partial_{\theta_{j}}-\partial_{\theta_{i}}\partial_{\bar{\theta}_{j}},$ and computing its commutator with \(q_i + q_j\), which yields
\begin{equation}
    \begin{aligned}\relax
        [\partial_{\perp i}\cdot \partial_{\perp j}, q_{i}+q_{j}] &= [\partial_{\perp i}\cdot \partial_{\perp j}, \ \x_{i}\partial_{\bar{\theta}_{i}}-\theta_{i}\partial_{\x_{i}}-\y_{i}\partial_{\theta_{i}}-\bar{\theta}_{i}\partial_{\y_{i}}]\\&\quad+[\partial_{\perp i}\cdot \partial_{\perp j},\ \x_{j}\partial_{\bar{\theta}_{j}}-\theta_{j}\partial_{\x_{j}}-\y_{j}\partial_{\theta_{j}}-\bar{\theta}_{j}\partial_{\y_{j}}]\\&
        =\partial_{\x_{j}}\partial_{\bar{\theta}_{i}}-\partial_{\bar{\theta}_{j}}\partial_{\x_{i}}-\partial_{\y_{j}}\partial_{\theta_{i}}+\partial_{\theta_{j}}\partial_{\y_{i}}\\&\quad+\partial_{\x_{i}}\partial_{\bar{\theta}_{j}}-\partial_{\bar{\theta}_{i}}\partial_{\x_{j}}-\partial_{\y_{i}}\partial_{\theta_{j}}+\partial_{\theta_{i}}\partial_{\y_{j}}= 0 \, .
    \end{aligned}
\end{equation} 
Using this result, the commutation with $q$ extends immediately to the full operator~\(P'_\perp\),
namely $P'_\perp q = q P'_\perp$. We then  obtain 
\begin{equation}
    P'_\perp q \,(\dots) \big|_{d-2}= q\, ( P'_\perp{\dots})\big|_{d-2} = 0 \, , 
\end{equation}
where in the last equality we used that 
the action of $q$ on any function in superspace is proportional to  $\x_i,\y_i,\theta_i,\thetab_i$ and thus vanishes on $\mathbb{R}^{d-2}$.
This shows that the second term in~\eqref{eq:almostThere} vanishes and thus completes the proof of~\eqref{eq:CoolestEq}. 

\subsection{A few examples}
Here we provide a couple of explicit $Q$-exact expressions. Consider the simplest combination,  $\mathcal{O} =(\partial_{\perp 1}\cdot\partial_{\perp 2})\Phi_{1}\Phi_{2}$.
In this case, we have $F_\Phi=\Phi_{1}\Phi_{2}$ and we wish to check that $\mathcal{O}$ and $Q((\partial_{\theta_{1}} \partial_{\x_2} + \partial_{\bar{\theta}_{1}} \partial_{\y_2}) F_\Phi)$ coincide when restricted to $d-2$ dimensions.  Since $Q$ acts independently on each superfield component, we need only compute the lowest component of $(\partial_{\theta_{1}} \partial_{\x_2} + \partial_{\bar{\theta}_{1}} \partial_{\y_2}) F_\Phi$, which is
\begin{equation}
    \partial_{\x_2} F_\Phi|_{\theta_{1}}+\partial_{\y_2} F_\Phi|_{\bar{\theta}_{1}}
    = \psi_{1}\partial_{\y_2}\varphi_{2}+\bar{\psi}_{1}\partial_{\x_2}\varphi_{2} \, .
\end{equation}
Thus, we conclude 
\begin{equation}
    \bigl((\partial_{\perp 1}\cdot\partial_{\perp 2})\Phi_{1}\Phi_{2}\bigr)\bigr|_{d-2}= Q\bigl((\psi_{1}\partial_{\y_2}\varphi_{2}+\bar{\psi}_{1}\partial_{\x_2}\varphi_{2} )|_{d-2}\bigr) \,.
\end{equation}
Notice that the proof in appendix~\ref{App:proof_Q_exact-subsec} used superspace.  Let us perform a sanity check by using variations of fields under the supersymmetry: one easily computes
\begin{equation}
\begin{aligned}
    \quad&\hspace{-1em}
Q\bigl((\psi_{1}\partial_{\y_{2}}\varphi_{2}+\bar{\psi}_{1}\partial_{\x_{2}}\varphi_{2})|_{d-2}\bigr)\\
&=\bigl({Q}\psi_{1}\partial_{\y_2}\varphi_{2}+{Q}\bar{\psi}_{1}\partial_{\x_2}\varphi_{2}-\psi_{1}\partial_{\y_2}{Q}\varphi_{2}-\bar{\psi}_{1}\partial_{\x_2}{Q}\varphi_{2}\bigr)|_{d-2}\\&=\bigl((\partial_{\y_{1}}\varphi_{1}+\y_{1}\omega_{1})\partial_{\y_{2}}\varphi_{2}+(\partial_{\x_{1}}\varphi_{1}+\x_{1}\omega_{1})\partial_{\x_{2}}\varphi_{2}\\&\quad-\psi_{1}\partial_{\y_{2}}(\x_{2}\psi_{2}-\y_{2}\bar{\psi}_{2})-\bar{\psi}_{1}\partial_{\x_2}(\x_{2}\psi_{2}-\y_{2}\bar{\psi}_{2})\bigr)|_{d-2}\\&=(\partial_{\y_{1}}\varphi_{1}\partial_{\y_{2}}\varphi_{2}+\partial_{\x_{1}}\varphi_{1}\partial_{\x_{2}}\varphi_{2}+\psi_{1}\bar{\psi}_{2}-\bar{\psi}_{1}\psi_{2})|_{d-2} \, , 
\end{aligned}
\end{equation}
and see that this indeed matches the left-hand side:
\begin{align}
    (\partial^{a_{1}}\Phi_{1}\partial_{a_{2}}\Phi_{2})|_{d-2}&=\bigl(\partial_{\x_{1}}\Phi_{1}\partial_{\x_{2}}\Phi_{2}+\partial_{\y_{1}}\Phi_{1}\partial_{\y_{2}}\Phi_{2}+\partial_{\bar{\theta}_{1}}\Phi_{1}\partial_{\theta_{2}}\Phi_{2}-\partial_{\theta_{1}}\Phi_{1}\partial_{\bar{\theta}_{2}}\Phi_{2}\bigr)|_{d-2}\\&=(\partial_{\y_{1}}\varphi_{1}\partial_{\y_{2}}\varphi_{2}+\partial_{\x_{1}}\varphi_{1}\partial_{\x_{2}}\varphi_{2}+\psi_{1}\bar{\psi}_{2}-\bar{\psi}_{1}\psi_{2})|_{d-2} \, .
\end{align}

For the other examples, these sorts of sanity checks are analogous but more lengthy. In what follows, we simply present the expressions without expliciting the checks. Consider, for example, $(\partial_{\perp 1}\cdot\partial_{\perp 2})^2\Phi_{1}\Phi_{2}$. We now have $P'_\perp=\partial_{\perp 1}\cdot\partial_{\perp 2}$ and $F_\Phi=\Phi_{1}\Phi_{2}$ hence
\begin{equation}
    \begin{aligned}
        \bigl(\partial_{\x_2} P'_\perp F_\Phi\bigr)\bigr|_{\theta_{1}}&=\partial_{\x_{2}}(\partial^{\ii_{1}}\Phi_{1}\partial_{\ii_{2}}\Phi_{2}+\partial_{\bar{\theta}_{1}}\Phi_{1}\partial_{\theta_{2}}\Phi_{2}-\partial_{\theta_{1}}\Phi_{1}\partial_{\bar{\theta}_{2}}\Phi_{2})|_{\theta_{1}}\\&=\partial^{\ii_{1}}\bar{\psi}_{1}\partial_{\ii_{2}}\partial_{\x_{2}}\varphi_{2}-\omega{}_{1}\partial_{\x_{2}}\bar{\psi}_{2}\, ,\\
        \bigl(\partial_{\y_2} P'_\perp F_\Phi\bigr)\bigr|_{\bar{\theta}_{1}}&=\partial_{\y_{2}}(\partial^{\ii_{1}}\Phi_{1}\partial_{\ii_{2}}\Phi_{2}+\partial_{\bar{\theta}_{1}}\Phi_{1}\partial_{\theta_{2}}\Phi_{2}-\partial_{\theta_{1}}\Phi_{1}\partial_{\bar{\theta}_{2}}\Phi_{2})|_{\bar{\theta}_{1}}\\&=\partial^{\ii_{1}}\psi_{1}\partial_{\ii_{2}}\partial_{\y_{2}}\varphi_{2}-\omega{}_{1}\partial_{\y_{2}}\psi_{2}\, ,
    \end{aligned}
\end{equation}
where we have introduced the Latin index notation $\ii \equiv \{d-1,d\}$. That is, $\partial_{\ii}=\{\partial_{\x}, \partial_{\y}\}$.
Therefore, 
\begin{equation}
  \begin{aligned}
\bigl((\partial_{\perp 1}\cdot\partial_{\perp 2})^2\Phi_{1}\Phi_{2}\bigr) \bigr|_{d-2} = Q\bigl((&\partial^{\ii_{1}}\bar{\psi}_{1}\partial_{\ii_{2}}\partial_{\x_{2}}\varphi_{2}-\omega{}_{1}\partial_{\x_{2}}\bar{\psi}_{2}
    \\ &+\partial^{\ii_{1}}\psi_{1}\partial_{\ii_{2}}\partial_{\y_{2}}\varphi_{2}-\omega{}_{1}\partial_{\y_{2}}\psi_{2})|_{d-2}\bigr) \, .
\end{aligned}
\end{equation}

As a final example, we consider three operators, $(\partial_{\perp 3}\cdot\partial_{\perp 1})(\partial_{\perp 2} \cdot \partial_{\perp 3}) \Phi_1 \Phi_2 \Phi_3$. 
This time, we have $P'_\perp F_\Phi=(\partial_{\perp 2} \cdot \partial_{\perp 3}) \Phi_1 \Phi_2 \Phi_3$, which gives
\begin{equation}
\begin{aligned}
    \bigl(\partial_{\x_1} P'_\perp F_\Phi\bigr)\bigr|_{\theta_3}+\bigl(\partial_{\y_1} P'_\perp F_\Phi\bigr)\bigr|_{\bar{\theta}_3} 
    & =\partial_{\y_1} \varphi_1 \partial^{\ii_2} \varphi_2 \partial_{\ii_3} \psi_3-\partial_{\y_1} \varphi_1 \psi_2 \omega_3 \\
    & \quad +\partial_{\x_1} \varphi_1 \partial^{\ii_2} \varphi_2 \partial_{\ii_3} \bar{\psi}_3-\partial_{\x_1} \varphi_1 \bar{\psi}_2 \omega_3 \, .
\end{aligned}
\end{equation}
That is, 
\begin{equation}
    \begin{aligned}
        \bigl((\partial_{\perp 3}\cdot\partial_{\perp 1})(\partial_{\perp 2}\cdot\partial_{\perp 3}) \Phi_1 \Phi_2 \Phi_3\bigr)\bigr|_{d-2} =
        Q\bigl(&(\partial_{\y_1} \varphi_1 \partial^{\ii_2} \varphi_2 \partial_{\ii_3} \psi_3-\partial_{\y_1} \varphi_1 \psi_2 \omega_3 \\
        & +\partial_{\x_1} \varphi_1 \partial^{\ii_2} \varphi_2 \partial_{\ii_3} \bar{\psi}_3-\partial_{\x_1} \varphi_1 \bar{\psi}_2 \omega_3)|_{d-2}\bigr) \, .
    \end{aligned}
\end{equation}

\section{Comments on localization}
\label{sec: App B}

\subsection{\(Q\)-invariance of the deformation term \texorpdfstring{$S_{\text{loc}}$}{S(loc)}}

In this appendix, we provide the proof that the localization term \eqref{S_loc_explicit}, by which we deform the path integral, is $Q$-closed.
Observe first that, for any pair of (odd) charges $Q_{\epsilon_1}$ and $Q_{\epsilon_2}$ and any fermionic field $\psi$ and bosonic field $\varphi'$, one has
\begin{equation}
    \begin{aligned}
    Q_{\epsilon_1}Q_{\epsilon_2}(\psi\varphi')
    & = Q_{\epsilon_1}(Q_{\epsilon_2}\psi\,\varphi'-\psi Q_{\epsilon_2}\varphi') \\
    & = (Q_{\epsilon_1} Q_{\epsilon_2} \psi) \varphi' + Q_{\epsilon_2}\psi Q_{\epsilon_1}\varphi' - Q_{\epsilon_1}\psi Q_{\epsilon_2}\varphi' + \psi\, Q_{\epsilon_1}Q_{\epsilon_2}\varphi' \,.
    \end{aligned}
\end{equation}
In the case $Q_{\epsilon_1}=Q_{\epsilon_2}$ the middle two terms cancel each other.
Thus, acting on \eqref{S_loc_explicit} by $Q$, we obtain
\begin{equation}
    Q^{2}\mathcal{V} = \int d^{d}x\, QQ(\psi Q\psi+\bar{\psi}Q\bar{\psi})
    = \int d^{d}x \, \bigl(Q^{2} \psi Q\psi+Q^{2}\bar{\psi}Q\bar{\psi}+\psi Q^{3}\psi+\bar{\psi}Q^{3}\bar{\psi}\bigr)\, .
    \label{eq:d^2V}
\end{equation}
Using the relations \eqref{QPhi and Q^2Phi} together with 
\begin{equation}
        Q^{3}\psi = -\partial_{\x}\varphi-\x\omega+m^{\x\y}(\partial_{\y}\varphi+\y\omega)\,, 
        \quad 
        Q^{3}\bar{\psi} = m^{\x\y}(\partial_{\x}\varphi+\x\omega)+\partial_{\y}\varphi+\y\omega\,,
\end{equation}
for each term in \eqref{eq:d^2V}, we obtain
\begin{equation}
    \begin{aligned}
        Q^{2}\psi Q\psi&=m^{\x\y}\psi\partial_{\y}\varphi+\y\omega m^{\x\y}\psi-{\color{blue!70!black}\bar{\psi}\partial_{\y}\varphi}-{\color{violet}\y\bar{\psi}\omega}\,,\\
        \psi Q^{3}\psi&=\psi m^{\x\y}\partial_{\y}\varphi+\psi m^{\x\y}(\y\omega)-{\color{purple}\psi\partial_{\x}\varphi}-{\color{teal}\x\psi\omega}\,,\\
        Q^{2}\bar{\psi}Q\bar{\psi}&=m^{\x\y}\bar{\psi}\partial_{\x}\varphi+\x\omega m^{\x\y}\bar{\psi}+{\color{purple}\psi\partial_{\x}\varphi}+{\color{teal}\x\psi\omega}\,,\\
        \bar{\psi}Q^{3}\bar{\psi}&=\bar{\psi}m^{\x\y}\partial_{\x}\varphi+\bar{\psi}m^{\x\y}(x\omega)+{\color{blue!70!black}\bar{\psi}\partial_{\y}\varphi}+{\color{violet}\y\bar{\psi}\omega}\,.
    \end{aligned}
\end{equation}
Upon summing these lines the colored terms cancel pairwise.  The remaining terms organize as total derivatives, which can be integrated by parts to give zero,
\begin{equation}
    \begin{aligned}
        & \int d^{d}x \, \bigl( Q^{2}\psi Q\psi + \psi Q^{3}\psi + Q^{2}\bar{\psi} Q\bar{\psi} + \bar{\psi} Q^{3}\bar{\psi} \bigr) \\
        & \qquad \qquad = \int d^{d}x \, m^{\x\y}\Bigl(\psi\partial_{\y}\varphi+\y\omega \psi+\bar{\psi}\partial_{\x}\varphi+\x\omega \bar{\psi}\Bigr)
        = 0 \,.
    \end{aligned}
\end{equation}
Thus, the localization term $Q\mathcal{V}$ is $Q$-closed.


\subsection{Deriving fermionic saddle point action}
\label{App:Fermionic_Saddle}
In this appendix, we show that the fermionic part of the SUSY action \eqref{eq:PS_action} can be written as \eqref{eq:S_d-1} when restricted to the localization locus fields \eqref{eq:psiS}. For the term involving the potential, we have 
\begin{align}
\psi_S \bar{\psi}_S V^{\prime \prime}(\varphi_S) & =
 (\sin \eta \bar{\xi}+\cos \eta \xi)(-\sin \eta \xi+\cos \eta \bar{\xi}) V^{\prime \prime}(\varphi_S) 
=\xi \bar{\xi} V^{\prime \prime}(\varphi_S)\,.
\end{align}
\\
For the kinetic term of the fermionic action $\partial^\mu\psi_S \partial_\mu\bar\psi_S$, we first split the derivatives into the parallel and the orthogonal parts. For the parallel part, we have
\begin{align}
    \partial^i\psi_S \partial_i\bar\psi_S = (\sin\eta \partial^i \bar{\xi}+\cos\eta \partial^i \xi)(-\sin\eta \partial_i \xi+\cos\eta \partial_i \bar{\xi})
    =  \partial^i \xi \partial_i  \bar{\xi} \, .
    \label{eq:||Part}
\end{align}
For the orthogonal part, we use $\partial_\x=\cos{\eta} \partial_r-\frac{\sin{\eta}}{r} \partial_\eta$ and $\partial_\y = \sin{\eta} \partial_r +\frac{\cos{\eta}}{r} \partial_\eta$. Then
\begin{align*}
    \partial^\perp\psi_S \partial_\perp\bar\psi_S  =&(\cos{\eta}\partial_{r}\psi_{S}-\frac{\sin{\eta}}{r}\partial_{\eta}\psi_{S})(\cos{\eta}\partial_{r}\bar{\psi}_{S}-\frac{\sin{\eta}}{r}\partial_{\eta}\bar{\psi}_{S})\\&+(\sin{\eta}\partial_{r}\psi_{S}+\frac{\cos{\eta}}{r}\partial_{\eta}\psi_{S})(\sin{\eta}\partial_{r}\bar{\psi}_{S}+\frac{\cos{\eta}}{r}\partial_{\eta}\bar{\psi}_{S})\, .
\end{align*}
The terms proportional to $\partial_r\psi_S\partial_r\bar\psi$  yield $\partial_r\xi\partial_r\bar\xi$. Moreover, the cross terms cancel, while  the terms with  $\partial_\eta\psi_S\partial_\eta\bar\psi_S$ give
\begin{equation}
    \frac{\mathrm{1}}{r^{2}}\partial_{\eta}\psi_{S}\partial_{\eta}\bar{\psi}_{S}=\frac{\mathrm{1}}{r^{2}}\partial_{\eta}(\sin{\eta}\bar{\xi}+\cos{\eta}\xi)\partial_{\eta}(-\sin{\eta}\xi+\cos{\eta}\bar{\xi})=\frac{\mathrm{\xi\bar{\xi}}}{r^{2}} \, .
\end{equation}
Gathering all the terms, we conclude
\begin{equation}
    S_{d-1}[\xi,\bar{\xi},\phi]
     = 2\pi \int d^{d-2}x_{\|} \int_0^{\infty} dr \, \Bigl( 
       \partial^{i}\xi\, \partial_{i}\bar{\xi}
       + \partial_{r}\xi\, \partial_{r}\bar{\xi}
       + \frac{\xi\bar{\xi}}{r^{2}}
       + \xi\bar{\xi}\, V''(\varphi_S) \Bigr) \,  , 
\end{equation}
as reported in \eqref{eq:S_d-1}.

\subsection{\(Q\)-invariance of saddle point action}
\label{app:Qinv_Fermionic_Saddle}

As a sanity check, in this section, we show the $Q$-invariance of the saddle point action given by
\begin{align}
	S[\varphi_S] &= \int d^{d}x\,\bigl( \partial^i \omega_S \partial_i \varphi_S+\omega_S V'(\varphi_S)+ \partial^\mu \psi_S \partial_\mu \bar{ \psi}_S+\psi_S \bar{\psi}_S V''(\varphi_S) \bigr) \,.
    \label{eq:sadAction}
\end{align}
where the bosonic part is in an already simplified form \eqref{eq:SofBosonSaddle}. 
Using the identities 
\begin{align}
    Q \varphi_S  =x \psi_S-y \bar{\psi}_S =r \xi \, , \qquad 
    Q \omega_S  =\partial_y \bar{\psi}_S-\partial_x \psi_S  =-\partial_r \xi\, ,
\end{align}
for the first term of the action, we obtain 
\begin{align*}
\begin{aligned}
    \int d^{d}x\,Q(\partial^i \omega_S \partial_i \varphi_S) & =\int d^{d}x\,\bigl(\partial^i(Q \omega_S) \partial_i \varphi_S+\partial^i \omega_S \partial_i(Q \varphi_S) \bigr)\\
    & =\int d^{d}x\,\bigl(-\partial^i(\partial_r \xi) \partial_i \phi-\partial^i\left(\frac{1}{r} \partial_r \phi\right) \partial_i(r \xi) \bigr) \\
    & =\int d^{d}x\,\bigl(\partial^i \xi \partial_i \partial_r \phi-\partial^i \partial_r \phi \partial_i \xi\bigr) =0 \, . 
\end{aligned}
\end{align*}
where, in the third line, we used integration by parts.  

For the second term of the action, we have
\begin{align}
    \int d^{d}x\, Q[\omega_S V^{\prime}\left(\varphi_S\right)]&=\int d^{d}x\,\bigl(V^{\prime}(\varphi_S) Q \omega_S+\omega_S Q V^{\prime}(\varphi_S)\bigr)\\ & =\int d^{d}x\,\bigl(-V^{\prime}(\varphi_S) \partial_r \xi+\omega_S V^{\prime \prime}(\varphi_S) Q \varphi_S \bigr) \\
    & =\int d^{d}x\,\bigl(-V^{\prime}(\varphi_S) \partial_r \xi-\partial_r V^{\prime}(\varphi_S) \xi \bigr) =0 \, ,
\end{align}
where in the third line we have used $\omega_S = -\partial_r\varphi_S$ and that $\partial_r\varphi_S V''(\varphi_S) = \partial_rV'$. 

Finally, for the fermionic part, using $Q \psi_S=Q\bar\psi_S = 0$, directly gives us that the kinetic term vanishes. For the term with the potential, we get
\begin{align}
    Q[\psi_S \bar{\psi}_S V^{\prime \prime}(\varphi_S)] & =\psi_S \bar{\psi}_S Q V^{\prime \prime}(\varphi_S) \\
    & =\psi_S \bar{\psi}_S V^{\prime \prime \prime}(\varphi_S) Q \varphi_S \\
    & =\psi_S \bar{\psi}_S V^{\prime \prime \prime}(\varphi_S)(x \psi_S-y \bar{\psi}_S) = 0 \, ,
\end{align}
since the fermions square to zero. This concludes the proof of $Q$-invariance of the saddle point action \eqref{eq:sadAction}. 
\subsection{Details on the one-loop determinant}
\label{sec:AppC}

In this appendix, we discuss some extra details about the one-loop determinant computation performed in \autoref{sec:1_loop}.

First, let us show how the action $S_{\text{loc}}^{(2)}$ defined in \eqref{S_loc} can be brought to the form given in \eqref{eq:SbandSf}. 
The fermionic part is easily obtained by rewriting \eqref{S_loc_explicit} in polar coordinates. Rewriting the bosonic part of \eqref{S_loc_explicit} in polar coordinates, we find
\begin{equation}
    S_b=\int d^{d-2}x_{\|} r d r d \eta\Bigl[(r \omega+\partial_r \varphi)^2+\frac{1}{r^2}\bigl(\partial_\eta \varphi\bigr)^2\Bigr] \, .
\end{equation}
Next, we diagonalize it. To this end, we integrate the last term by parts and perform the field redefinitions $\varphi / r \rightarrow \varphi$ and $\omega r+\partial_r \varphi \rightarrow \omega$, which has a trivial Jacobian. 
In terms of these new fields, we obtain
\begin{equation}
    S_b=\int d^{d-2}x_{\|} r d r d \eta\left(\omega^2-\varphi \partial_\eta^2 \varphi\right)\, ,
\end{equation}
which matches the expression in \eqref{eq:SbandSf}.

Let us also mention in more detail how to perform the Gaussian integrals which give the result in \eqref{Z_1loop=1}.
As explained in the main text, the actions $S_b$ and $S_f$ can be written in the Fourier modes \eqref{eq:fermFourier} as written in \eqref{Sb_Sf_fourire}. The result can be also expressed in a matrix form as follows
\begin{align}
   & S_b=2 \pi \int d^{d-2} x_\| \int r d r  \sum_{k=1}^{\infty} \left(\omega_{-k}, \omega_{k}, \varphi_{-k} ,\varphi_{k}\hspace{0.1cm}\right) 
   {
    \def\arraycolsep{2pt}
    \renewcommand\arraystretch{0.7}
    \begin{pmatrix}
    0 & 1 & 0 & 0 \\
    1 & 0 & 0 & 0 \\
    0 & 0 & 0 & k^2 \\
    0 & 0 & k^2 & 0
    \end{pmatrix}
   \begin{pmatrix}
    \omega_{-k} \\
    \omega_{k} \\
    \varphi_{-k} \\
    \varphi_{k}
    \end{pmatrix}
    }
     \, ,
      \nonumber
\\
    &S_f= 2 \pi \int d^{d-2} x_\| \int r d r  \left[\sum_{\myatop{k=0} {k\neq 1}}^{\infty} 
    {
    \def\arraycolsep{0.5pt}
    \renewcommand\arraystretch{0.7}
 \begin{pmatrix} \psi_{-k},
 \psi_{k},
 \bar{\psi}_{-k},
 \bar{\psi}_{k} \end{pmatrix}
    \begin{pmatrix}
    0 & i k & 0 & -1 \\
    -i k & 0 & -1 & 0 \\
    0 & 1 & 0 & i k \\
    1 & 0 & -i k & 0
     \end{pmatrix} 
   \begin{pmatrix}
    \psi_{-k} \\
    \psi_{k} \\
    \bar{\psi}_{-k} \\
    \bar{\psi}_{k}
    \end{pmatrix}
    }
    +4 \alpha^{+}_{-1} \alpha^{+}_1 \right]
    \, .
    \nonumber
\end{align}
We notice that the determinant of the bosonic matrix is $k^4$, while the Pfaffian of the fermionic matrix is $|k^2-1|$. The bosonic and fermionic Gaussian integrals are respectively normalized as
\begin{equation}
\label{Def:Gaussian_integrals}
     \int \prod_{i=1}^n dx_i e^{-\frac{1}{2} x_i M_{ij} x_j}= \frac{(2\pi)^{n/2}}{\sqrt{\det M}}
      \, , 
    \qquad
        \int \prod_{i=1}^n d\theta_i e^{-\frac{1}{2} \theta_i A_{ij} \theta_j}= \mbox{Pf}(A)
     \, .
\end{equation}
The Pfaffian divided by the square root of the determinant gives $|k^2-1|/k^2$ and it is multiplied by a factor of $2$  from the contribution of $ \alpha^{+}_{-1} \alpha^{+}_1$; altogether these terms give the expression inside the brackets of \eqref{Z_1loop=1}.

In principle, there would be a discrepancy of a factor of $\sqrt{2\pi}$ for each bosonic mode because of the normalization of \eqref{Def:Gaussian_integrals}.
For this reason we believe that it is more appropriate to use the fermionic measure $[\mathcal{D}\psi][\mathcal{D}\psib]$, which we interpret as a path integral generalization of \eqref{fermionic_measure}, which effectively weights each fermionic mode by $1/\sqrt{2\pi}$, thus producing the final result of \eqref{Z_1loop=1}.


\section{Comments on the interpolation argument: an explicit example}

In \autoref{sec:gen_Cardy}, we provided a rather abstract algorithm to prove dimensional reduction of generic theories. 
This was based on a split of the action into $S_\parallel$ and $S_\perp$. 
In the following, we give a further example of this split, and we also explicitly show that $S_\perp$ is $Q$-exact.
To this end, let us consider a higher derivative generalization of the kinetic term of  \eqref{eq:PS_action_superfield}
\begin{equation}
    S_{\text{HD}}=\int d^d x d \bar{\theta} d \theta\left[-\frac{1}{2} \Phi \square^n \Phi\right], \qquad \square^n=\left(\partial^\mu \partial_\mu\right)^n+2 n\left(\partial^\mu \partial_\mu\right)^{n-1} \partial_{\bar{\theta}} \partial_\theta \, .
\end{equation}
For future convenience, we split the operator $\Box^n$ into parallel and orthogonal parts, and use the binomial expansion. This gives
\begin{align}
    \square^{n}=(\partial_{\|}^{2})^{n}+A+2nB\partial_{\bar{\theta}}\partial_{\theta} \, ,
\end{align}
where we have defined 
\begin{equation}
    A\equiv \sum_{k=0}^{n-1} C_{k+1}^n(\partial_{\|}^{2})^{n-k-1}(\partial^{\ii} \partial_{\ii})^{k+1}, \quad B\equiv\sum_{k=0}^{n-1} C_k^{n-1}(\partial_{\|}^{2})^{n-1-k}(\partial^{\ii} \partial_{\ii})^k\,.
\end{equation}
Here $C_{k}^n$ are standard binomial coefficients:
\begin{equation}
    C_{k}^n = \frac{n!}{k!(n-k)!} \, ,
\end{equation}
and we have introduced the Latin index ($\ii \text{ or } \jj$) notation $\ii \equiv \{d-1,d\}$, hence $x^{\ii}=\{\x, \y\}$.

Following the prescription of  \autoref{sec:gen_Cardy}, we split the action $S_{\text{HD}}$ into parallel and orthogonal parts,
\begin{align}
     S^{\parallel}_{\text{HD}}
=\int d^{d}x \bigl( - \varphi(\partial_{\|}^{2})^{n}\omega-\psi(\partial_{\|}^{2})^{n} \bar{\psi} \bigr) \, ,
     \quad
    S^{\perp}_{\text{HD}}
    =\int d^{d}x \bigl( -\varphi A\omega-\psi A\bar{\psi}-n\omega B\omega \bigr) \, .
\end{align}
In  $Q$-cohomology, $S^{\parallel}_{\text{HD}}$ is equal to the reduced action $ S^{d-2}_{\text{HD}}
=\int d^{d-2}x_{\|} \bigl( -\frac{1}{2} \varphi(\partial_{\|}^{2})^{n}\varphi \bigr)$  by the argument of \autoref{sec:gen_Cardy}.
In the main text, we also showed that for scalar theories $S^{\perp}_{\text{HD}}$ must be $Q$-exact. In the following, we check that this statement is true for $S^{\perp}_{\text{HD}}$ and we provide the explicit form of the $Q$-exact term.
Inspired by the $Q$-exact form \eqref{eq:bosoSector} of the standard orthogonal action, our guess for $S^{\perp}_{\text{HD}}$ is given by
\begin{equation}
    {S^\perp_{\text{HD}} = Q\int d^{d}x\,\omega C \psi_\perp\, ,} \qquad C \equiv \sum_{k=0}^{n-1} C_{k+1}^n(\partial_{\|}^{2})^{n-k-1}(\partial^{\ii} \partial_{\ii})^{k} \, ,
    \label{eq:SHDQexact}
\end{equation}
where $\psi_\perp$ was defined in \eqref{psi-par-perp}. Notice that the operator $C$ is related to $A$ as follows: $C\partial^{\ii}\partial_{\ii} = A$. 

Let us check \eqref{eq:SHDQexact}. Using the relations in \eqref{QPhi and Q^2Phi}, \eqref{Qpsi-par-perp}, and that $C\partial^{\ii}\partial_{\ii} = A$, we obtain
\begin{align}
    \nonumber
    Q\int d^{d}x\, \omega C \psi_\perp &= \int d^{d}x\, \Bigl( -\omega C \partial^{\ii} \partial_{\ii}  \varphi- 2 \omega C\omega - \omega C( r \partial_r\omega) \\[-1ex]
    &\qquad\qquad\quad + \partial_{\x}\psi
   C \partial_{\x}\bar{\psi}-\partial_{\y}\bar{\psi} C \partial_{\y}\psi
   +\partial_{\x}\psi C \partial_{\y}\psi-\partial_{\y}\bar{\psi}C\partial_{\x}\bar{\psi}\Bigr) 
   \label{eq:fermPart}\\
    &= S^\perp_{\text{HD}} - \int d^{d}x \,\bigl(2 \omega C\omega + \tfrac{1}{2} \omega C(r \partial_r\omega) + \tfrac{1}{2} C\omega \, r\partial_r \omega - n\omega B\omega\bigr) \,.
    \label{eq:NotYetDone}
\end{align}
Here we have integrated by parts $-\omega C \partial^{\ii} \partial_{\ii} \varphi + \partial_{\x}\psi C \partial_{\x}\bar{\psi}-\partial_{\y}\bar{\psi} C \partial_{\y}\psi$ to reproduce parts of~$S^\perp_{\text{HD}}$, and to split $\omega C(r \partial_r\omega)$ into two terms that will be more convenient momentarily.  The third fermionic term $\partial_{\x}\psi C \partial_{\y}\psi$ can be integrated by parts into either one of the forms $-(\partial_{\y}C\partial_{\x}\psi)\psi$ and $-\psi(\partial_{\x}C\partial_{\y})\psi$, but these two forms are opposite since $\partial_{\y}C\partial_{\x}\psi=\partial_{\x}C\partial_{\y}\psi$, so their (equal and opposite) integrals vanish.  The last term $-\partial_{\y}\bar{\psi}C\partial_{\x}\bar{\psi}$ has a vanishing integral for the same reason.

To finish proving that our guess is correct, we show that the extra contribution in~\eqref{eq:NotYetDone} vanishes.
Observe first that $[r\partial_r, \partial^{\ii}\partial_{\ii}] = - 2 \partial^{\ii}\partial_{\ii}$, from which we deduce the commutator
\begin{equation}
    [r\partial_r, C] - 2 C = \sum_{k=1}^{n-1} C^{n}_{k+1} (\partial_{\|}^{2})^{n-k-1} (\partial^{\ii} \partial_{\ii})^k (-2k-2)
    = - 2 n B
\end{equation}
using that $(-2k-2)C^{n}_{k+1}=-2n C^{n-1}_{k}$.  The (integrand of the) extra term in~\eqref{eq:NotYetDone} is thus
\begin{equation}
      2 \omega C\omega + \tfrac{1}{2} \omega C( r \partial_r\omega) + \tfrac{1}{2} C \omega \, r\partial_r \omega - n\omega B\omega
      = \omega C\omega + \tfrac{1}{2} C \omega \, r\partial_r \omega + \tfrac{1}{2} \omega \, r\partial_r C\omega
      = \tfrac{1}{2} \partial_{\ii} \bigl( x^{\ii} \omega C\omega \bigr) ,
\end{equation}
which is a total derivative.  This proves~\eqref{eq:SHDQexact}.

\bibliography{biblio}
\bibliographystyle{JHEP}

\end{document}